\documentclass[amsmath,amssymb,aps,prb,fleqn,twocolumn,showpacs]{revtex4}

\newcommand{\bra}[1]{\ensuremath{\langle#1|}}
\newcommand{\ket}[1]{\ensuremath{|#1\rangle}}
\newcommand{\bracket}[2]{\ensuremath{\langle#1|#2\rangle}}

\newcommand{\avg}[1]{\ensuremath{\langle #1 \rangle}}
\newcommand{\op}[1]{\ensuremath{\hat{#1}}}

\usepackage{dsfont}
\usepackage{alltt}

\usepackage{ifpdf}
\ifpdf
    \usepackage{psfrag}
    \usepackage[pdftex]{epsfig}
\else
    \usepackage{psfrag}
    \usepackage[dvips]{epsfig}
\fi

\begin{document}

\title{Minimally Entangled Typical Thermal State Algorithms}
\date{\today}
\author{E.M. Stoudenmire}
\affiliation{Department of Physics, University of California, Santa Barbara, CA 93106-4030}
\author{Steven R. White}
\affiliation{Department of Physics and Astronomy, University of California, Irvine, CA 92697}



%
\begin{abstract}
We discuss a method based on sampling minimally entangled typical thermal states (METTS) that can
simulate finite temperature quantum systems with a computational cost comparable to ground state DMRG. 
Detailed implementations of each step of the method are presented, along with efficient algorithms for working with 
matrix product states and matrix product operators.
We furthermore explore how properties of METTS can reveal characteristic order and excitations of systems and discuss why 
METTS form an efficient basis for sampling. 
Finally, we explore the extent to which the average entanglement of a METTS ensemble is minimal.  
\end{abstract}

\maketitle

\section{Introduction}

According to the elementary principles of quantum statistical mechanics, the average of an observable is found by computing 
\begin{equation}
\langle A \rangle = \mbox{Tr}[\rho\,A] = \frac{1}{\mathcal{Z}} \mbox{Tr}[e^{-\beta H}A] \: . \label{eqn:trace}
\end{equation}
While performing such a calculation directly is usually intractable for quantum many body systems,
one can still approximate the expectation value of $A$ by strategies based on sampling. For instance, 
many Quantum Monte Carlo methods take advantage of the Trotter decomposition of $\rho$ to express an
expectation value as a sum over virtual classical paths, which can then be numerically sampled. In such
approaches, the act of randomly choosing a path corresponds to sampling both
the thermal and quantum fluctations together.

However, it is also possible to sample the thermal fluctuations only, randomly selecting
quantum states rather than classical ones. This may be done by expanding the trace in Eq.~(\ref{eqn:trace}) in
terms of an orthonormal basis $\ket{i}$ such that the expectation value of $A$ takes the form
\begin{eqnarray}
\langle A \rangle & = &  \frac{1}{\mathcal{Z}} \sum_i \bra{i} e^{-\beta H/2}\, A\  e^{-\beta H/2} \ket{i} \nonumber \\
& = & \frac{1}{\mathcal{Z}} \sum_i P(i) \bra{\phi(i)} A\, \ket{\phi(i)} \: .\label{eqn:thermal_avg}
\end{eqnarray}
In the first line we have used the cyclic property of the trace and in the second line
the set of normalized states $\ket{\phi(i)}$ are defined as 
\begin{equation}
\ket{\phi(i)} = P(i)^{-1/2} e^{-\beta H/2} \ket{i} \label{eqn:phi_state}
\end{equation}
together with the (unnormalized) probability distribution
\begin{equation}
P(i) = \bra{i}e^{-\beta H}\ket{i} \ . \label{eqn:weight}
\end{equation}
One may then estimate $\avg{A}$ by sampling the states $\ket{\phi(i)}$ with probability $P(i)/\mathcal{Z}$
and averaging the expectation values $\bra{\phi(i)}A\ket{\phi(i)}$ computed at each step.

Though such a procedure could be carried out using any orthonormal basis of states $\ket{i}$, not every choice would lead
to an efficient algorithm. In particular, the computational cost of working with a quantum state goes up the more
entangled it is, as measured by the von Neumann entanglement entropy across bipartitions of the system.


Therefore a natural choice for the basis $\ket{i}$ is the set of classical product states (CPS). 
These are states with wavefunctions of the form
\begin{equation}
\ket{i} = \ket{i_1}\ket{i_2}\ket{i_3}\ldots\ket{i_N} \:,
\end{equation}
where the $i_j$ label states in a local basis that may be chosen arbitrarily for each site $j$.
As a result of their product structure, CPS have an entanglement entropy that is exactly zero.

One therefore might expect that, of all ensembles of states $\ket{\phi(i)}$,
those produced from CPS have the least entanglement.
Moreover, such $\ket{\phi(i)}$ have all of the properties one intuitively expects of 
typical states of a thermal system. 
At high temperature they are effectively classical, while at lower temperatures 
they spontaneously break symmetries of the Hamiltonian in systems that exhibit long range order. 
They do not exhibit unphysical non-local correlations, and remain factorized over parts of the system which do not interact. 
And, expectations of observables $\bra{\phi(i)}A\ket{\phi(i)}$ lie very close to the thermal average.
For such reasons as well as others that we will discuss below, these states have therefore been 
designated minimally entangled typical thermal states, or METTS.\cite{white-09}

Not only does each METTS provide a good characterization of the thermal ensemble,
but there exists a simple algorithm, the pure state algorithm, for sampling many of them efficiently. 
This is acheived by a random walk through the set of METTS where the next state is constructed from a CPS obtained
by measuring, or collapsing, the previous one. 
This collapsing step automatically ensures that the METTS are sampled with the 
correct distribution, and it allows one to optimize the the sampling process by an appropriate choice of basis. 
Moreover, collapsing each state may be seen as sampling the quantum fluctuations of the system, a fact that 
can be used to accelerate the calculation of certain physical observables. 

In this paper, we elaborate upon the implementation of the pure state algorithm for sampling METTS and its usage in realistic simulations. 
First, in section \ref{sec:pure_state_method} we provide a more detailed explanation of the pure state algorithm and its
implementation. We begin with a discussion of various methods for carrying out the imaginary time evolution implicit in Eq.~(\ref{eqn:phi_state}), 
as it is the most computationally intensive part of the algorithm and the step where the most significant approximations must be made. 
Then, we turn to the measurement of physical observables. While it is most convenient to calculate quantities like the total energy using matrix product
operators (MPOs), we will show how local observables can be calculated more efficiently by taking advantage of the redundancy inherent in the matrix product state 
(MPS) formalism. We then conclude the section by discussing an efficient method for collapsing a pure state into a CPS. 
The basis into which the CPS is measured can be arbitrary, and we will demonstrate how this freedom may be exploited both to improve the sampling autocorrelation
time and to help calculate challenging observables. 


In section \ref{sec:properties} we turn our focus to the properties of the METTS themselves. Properties of individual METTS can provide 
insight into the order that is present in a system as well as characteristic thermal fluctuations. On the other hand, studying the entire METTS ensemble
will show us why it is such an efficient basis for sampling.  Finally, we quantify the sense in which METTS are minimally entangled
and conjecture that although there exist thermal decompositions with less entanglement than METTS at some temperatures, no decomposition can outperform
an optimal METTS ensemble at all temperatures.

Throughout, whenever we find it helpful to show the results of an explicit calculation, we will choose as our model of interest the $S=1$ bilinear-biquadratic chain with 
Hamiltonian
\begin{equation}
H = \cos\theta \sum_i \vec{S}_i\cdot\vec{S}_{i+1} + \sin\theta \sum_i (\vec{S}_i\cdot\vec{S}_{i+1})^2 \ . \label{eqn:S1Ham}
\end{equation}
In particular we will focus on two special cases of this model, namely \mbox{$\theta = 0$}, the $S=1$ Heisenberg
antiferromagnet and \mbox{$\theta = \arctan[1/3]$}, the AKLT model posessing an $S=1$ valence bond solid ground state.~\cite{aklt}
At \mbox{$T=0$}, both models are in the same phase, the Haldane phase, which is characterized by, among other things, 
a gap to all excitations\cite{haldane-83} and a doubly degenerate entanglement spectrum.~\cite{pollmann-09}

%
%

\section{Producing METTS With the Pure State Method \label{sec:pure_state_method}}

The pure state method consists of the following simple algorithm for producing a series of METTS such that they are correctly distributed with probability $P(i)/\mathcal{Z}$:

\begin{enumerate}
\item Choose a CPS $\ket{i}$.
\item Compute the METTS $\ket{\phi(i)} = e^{-\beta H/2} \ket{i} P(i)^{-1/2} $ and calculate observables of interest.
\item Collapse a new CPS $\ket{i^\prime}$ from $\ket{\phi(i)}$ with probability $p(i\rightarrow i^\prime) = |\bracket{i^\prime}{\phi(i)}|^2$ and return to step 2.
\end{enumerate}
We will shortly discuss how to carry out these steps in practice. However, let us first 
see why producing METTS in this way is guaranteed to give the correct distribution. 

Consider the ensemble of all CPS $\ket{i}$ initially distributed with probability $P(i)/\mathcal{Z}$. If we randomly choose a CPS $\ket{i}$ from this ensemble and then follow the steps above, the probability of measuring a particular CPS $\ket{j}$ at the end of step 3 will be
\begin{align}
 \sum_i \frac{P(i)}{\mathcal{Z}}\:p(i\rightarrow j)  & = \sum_i \frac{P(i)}{\mathcal{Z}}\ |\bracket{j}{\phi(i)}|^2 \nonumber \\
& = \sum_i \frac{P(i)}{\mathcal{Z}}\: \frac{|\bra{j} e^{-\beta H/2} \ket{i}|^2}{P(i)} \nonumber \\
& = \bra{j}e^{-\beta H}\ket{j}/\mathcal{Z} \nonumber \\
& = P(j)/\mathcal{Z} \ .
\end{align}
That the resulting ensemble of all such $\ket{j}$ have the same distribution as the original ensemble is a consequence of the detailed balance condition 
\begin{equation}
\frac{P(i)}{\mathcal{Z}}\: p(i\rightarrow j)= \frac{P(j)}{\mathcal{Z}}\: p(j \rightarrow i) \ .
\end{equation}
Therefore the ensemble of CPS with distribution $P(i)/\mathcal{Z}$ is a fixed point of this process.
Finally then, because each METTS $\ket{\phi(i)}$ is produced deterministically from a CPS $\ket{i}$ in step 2, they will be generated with probability $P(i)/\mathcal{Z}$ as well.

In addition to generating METTS with the correct distribution, the pure state method has other important properties that make it advantageous for performing simulations. 
First, it may be defined in a completely general way, allowing one to choose a specific implementation based on the problem of interest. 
Next, it does not rely on a rejection method to perform sampling. That is, every METTS produced (in what is the costliest step of the algorithm) can be used to perform measurements. 
And, the algorithm is readily parallelizable since its sampling method is that of a Markovian random walk.

\begin{figure}[htp]
\includegraphics[width=\columnwidth]{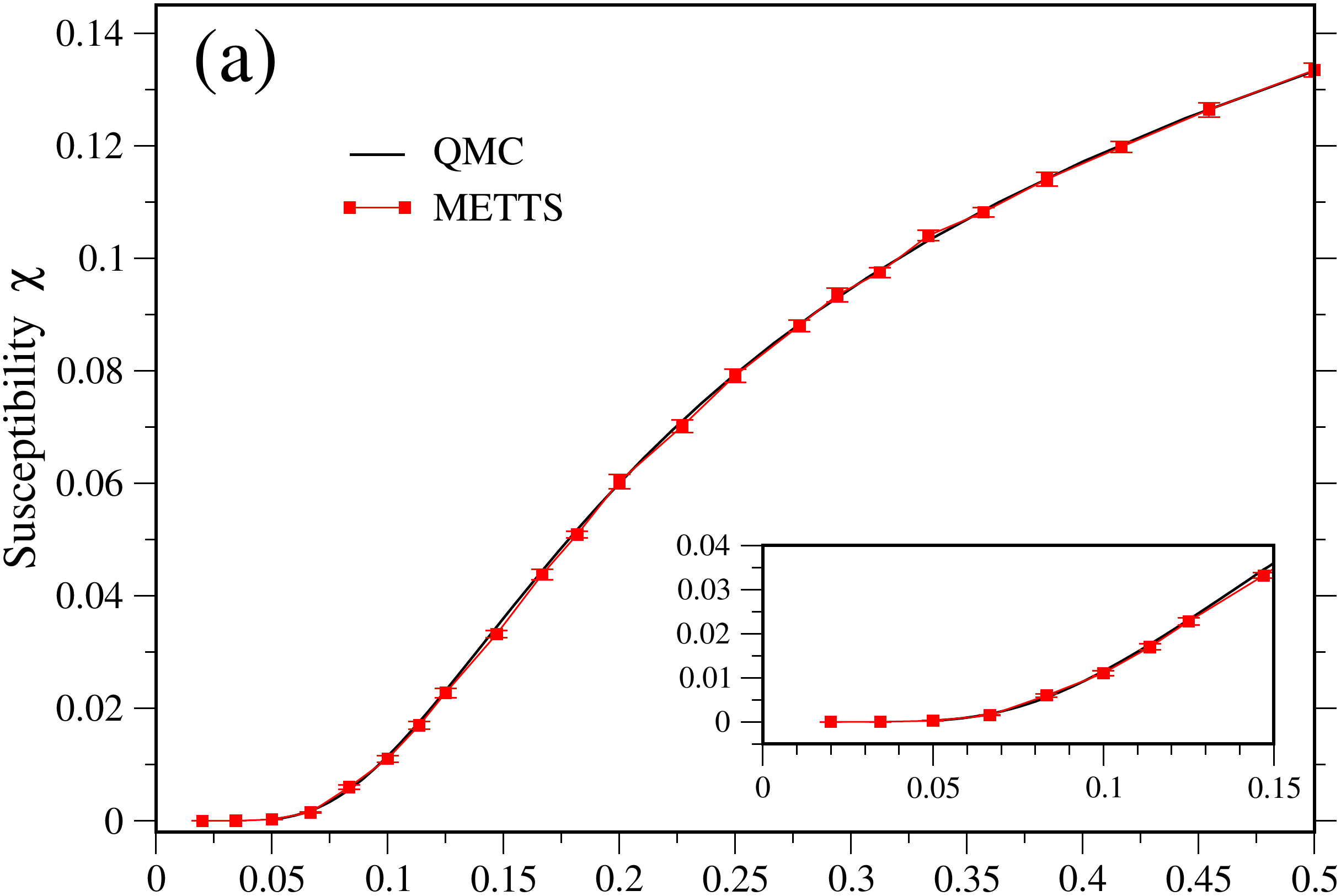} 
\vspace{0.0cm} \\
\includegraphics[width=\columnwidth]{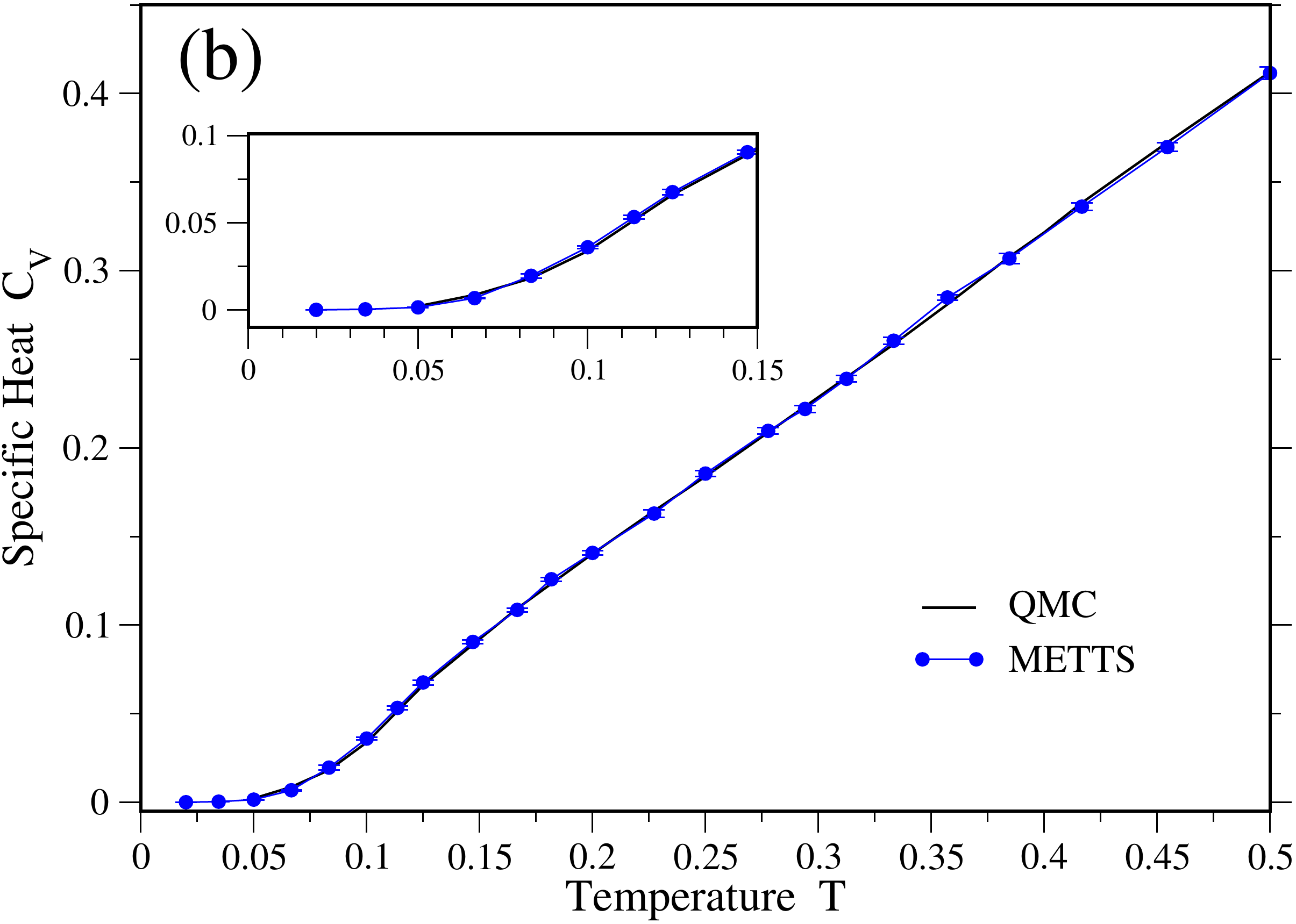}
\caption{Low temperature measurements of the (a) susceptibility and (b) specific heat of the $S=1$ Heisenberg chain with $N=100$ sites. }
\label{fig:lowT}
\end{figure}

Most importantly, the computational cost of the pure state algorithm is only modestly greater than that of ground state DMRG. 
The typical bond dimension $m$ of the matrix product state representing each METTS produced by the imaginary time evolution step ranges from \mbox{$m=1$} for \mbox{$\beta=0$} to \mbox{$m=m_0$} for very large $\beta$, where $m_0$ is the bond dimension necessary
to represent the ground state using DMRG. And, as shown below, the imaginary time evolution for a system with $N$ sites scales as $N m^3$ per timestep. 
Therefore the cost of producing each METTS scales as $N m^3 \beta$ (ground state DMRG scales as $Nm^3$). 
The CPS collapse step is much less costly, scaling only as as $N m^2$. 

We have therefore been able to reach significantly lower temperatures with the METTS approach than with previous DMRG-based finite temperature methods. 
For example, Feiguin and White were able to simulate the $S=1$ Heisenberg chain down to about $T=0.05$ using the ancilla method, where about $m=500$
DMRG states had to be kept at the lowest temperatures when using a truncation error cutoff of $1\!\times\!10^{-10}$. This is because their matrix product states had to encode both the system and environment spins.\cite{feiguin-ancilla}

In contrast, to reach comparable temperatures using METTS requires only about \mbox{$m=60$} states when using the same cutoff. 
This has allowed us to produce accurate results down to at least \mbox{$T=0.02$}, as shown in Fig.~\ref{fig:lowT}. 

And, since the computational cost of METTS is comparable to that of ground state DMRG, we expect that it can treat comparably 
large ladders and two-dimensional systems. This promises to be of great value for studying frustrated and fermionic models beyond one dimension ---
especially models with non-trivial phases at finite temperature. 

Having defined the pure state method in a general setting, let us look more closely at each step. 
We will discuss how to produce and sample METTS in detail and
demonstrate that there is great flexibility within the basic algorithm that can allow us to optimize it for our system of interest. 


\subsection{Imaginary Time Evolution}

Since the imaginary time evolution is both the costliest step of the algorithm 
and the source of the most numerical error, we will begin by discussing some of the best methods for evolving a state in imaginary time
and their relative advantages and disadvantages with regards to producing METTS. 

In what follows, we will find it convenient to represent wavefunctions as
matrix product states (MPS) and, in certain cases, operators as matrix product operators (MPO). 
A brief review of this formalism may be found in the Appendix.

\subsubsection{Trotter-Suzuki Method}

For a one dimensional system, the simplest way to implement time evolution is to use a Trotter-Suzuki decomposition of the time evolution operator, evolving each bond one at a time. 
When using this method for our 1D benchmark calculations, we chose to do a second-order breakup as in the 
time-dependent DRMG method of White and Feiguin\cite{white-04}.
Assuming that the Hamiltonian consists of a sum of nearest-neighbor bond hamiltonians $H = \sum_n H_{n}$, decompose $e^{-\tau H}$ as
\begin{equation}
e^{-\tau H}\! \simeq e^{-\tau H_1/2} e^{-\tau H_2/2} \ldots e^{-\tau H_2/2} e^{-\tau H_1/2} + \mathcal{O}(\tau^3) \, . \label{eqn:TSbreakup}
\end{equation}

We emphasize here that the labeling of the terms in the Hamiltonian is arbitrary. However, for a 1D system with nearest-neighbor interactions 
it is convenient to let $H_n$ refer to the interactions on the $n^{\mbox{\tiny th}}$ bond, making the operator decomposition above perfectly suited for a single DMRG sweep.
Because each factor, or `gate', in this breakup is a local operator, it can be computed exactly such that the only error in this treatment of the time evolution 
operator is the finite time step Trotter error. 
Then, each gate may be applied directly to the MPS representing the state of the system by using DMRG to expose the two sites around the bond to be time evolved.

\begin{figure}[htp]
\includegraphics[width=\columnwidth]{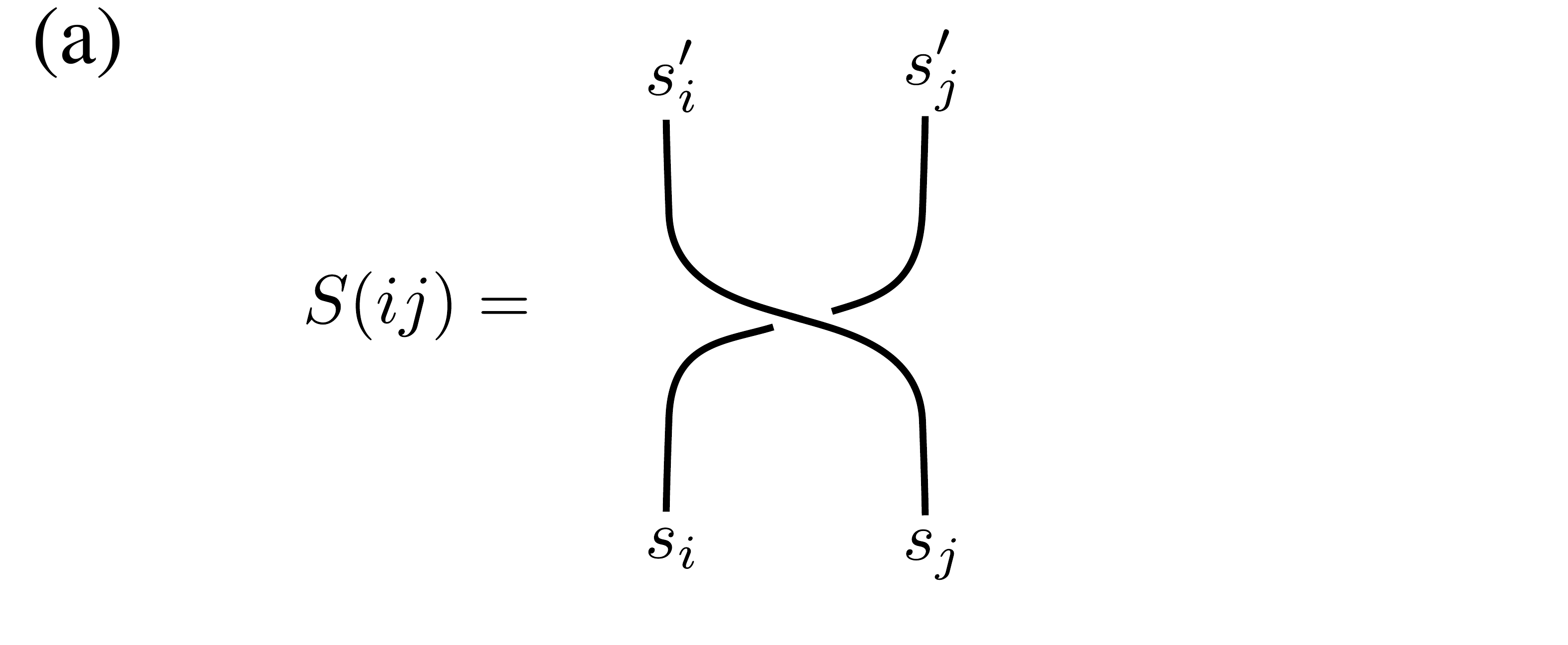}
\includegraphics[width=\columnwidth]{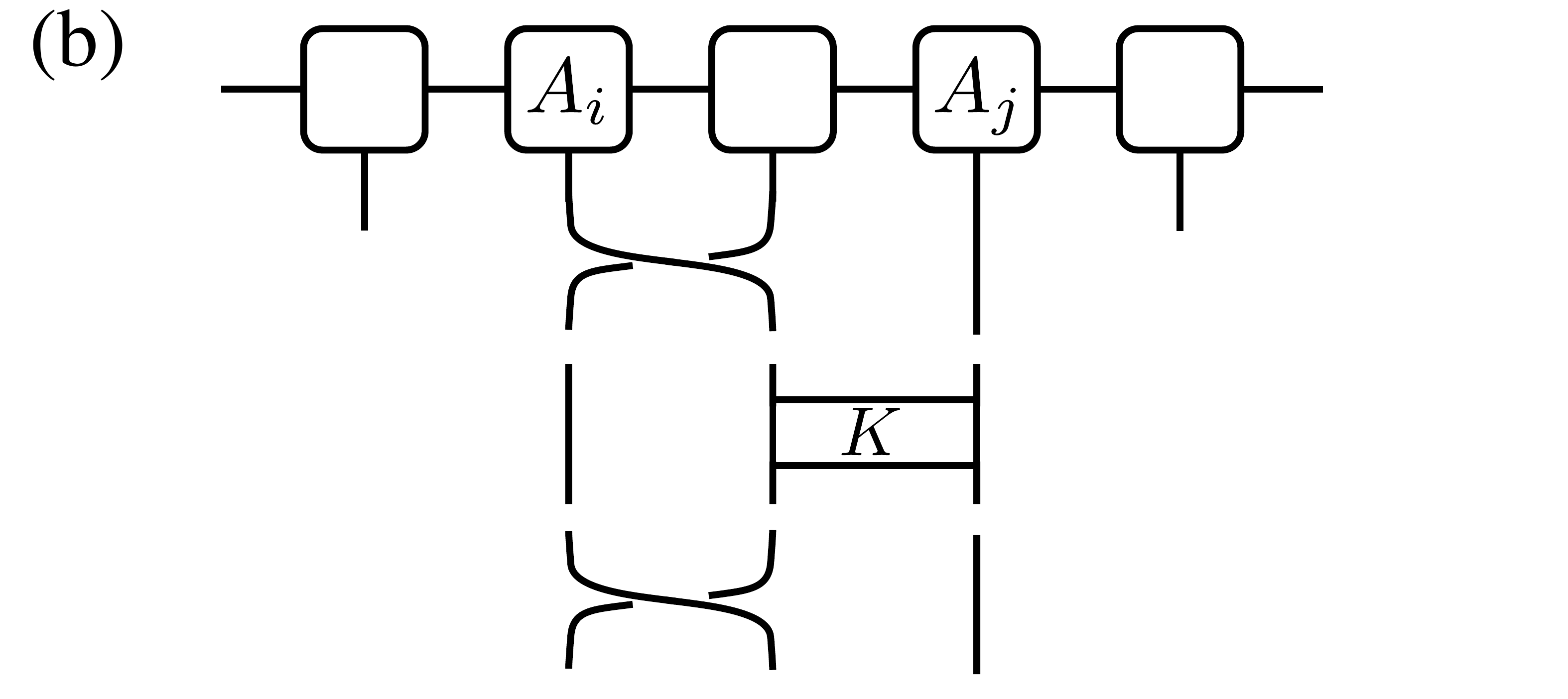}
\caption{Diagrams illustrating (a) the structure of a swap gate tensor, which is a product of two identity tensors and exchanges the states of two lattice sites
and (b) the application of swap gates to an MPS, allowing a Trotter time evolution gate to act on effective nearest-neighbor sites while actually evolving 
a system with longer range interactions.}
\label{fig:swapgate}
\end{figure}

While the Trotter-Suzuki method is accurate and convenient, it is only directly applicable to systems with nearest-neighbor interactions along the path taken
by the MPS through the sites of the lattice. This makes it hard to apply to 2D systems, including those with nearest-neighbor interactions.
Even quasi-1D ladders and 1D chains with longer range interactions cannot be treated by the above method without some modifications. 

Various methods have therefore been developed to extend Trotter time evolution to such systems, for example Feiguin and White's method based on
a Runge-Kutta differential equation solver.\cite{feiguin-timestep} However, such methods turn out to be too inefficient for our purposes. 
So, we will discuss some alternative methods below that are more effective for these challenging systems.

\subsubsection{Trotter-Suzuki with Swap Gates}

The first method extends the Trotter-Suzuki method by utilizing swap gates, a tool that is familiar in quantum computing. A swap gate $S(ij)$ exchanges 
the states on two identical sites $i$ and $j$. For a spin or boson system, \mbox{$S(ij) = \delta_{s_i s_j^\prime} \delta_{s_j s_i^\prime}\,$}; the diagrammatic representation
of which is shown in Fig.~\ref{fig:swapgate}a. Swap gates can be used for fermionic systems as well. In the case of spinless fermions, for instance, $S(ij)$ takes the same form as for bosons, but with an overall minus sign multiplying the case in which both sites are occupied.

Now, a generic Trotter-Suzuki breakup of the time evolution operator for a Hamiltonian containing only one site or two site interaction terms can be written in the form
\begin{equation}
e^{-\tau H} \simeq K_{i_1j_1} K_{i_2j_2} K_{i_3j_3} \ldots
\end{equation}
were the sites acted on by a given gate $K_{ij}$ are not necessarily nearest-neighbors along the path taken by the MPS through the lattice. 
However, $K_{ij}$ can be written in terms of a gate that does act on neighboring sites by using the identity
\begin{equation}
K_{ij} = \Theta_{i\, j-1} K^{(ij)}_{j-1\,j} \Theta_{j-1\, i}
\end{equation}
where the swap operators $\Theta$ can be defined as a product of nearest-neighbor swap gates as
\begin{align}
\Theta_{i\, j-1} & = S_{i\, i+1}S_{i+1\, i+2}\ldots S_{j-2\,j-1} \\ 
\Theta_{j-1\, i} &= S_{j-2\,j-1} \ldots S_{i+1\, i+2}S_{i\, i+1}  = \Theta_{i\,j-1}^\dagger \ . \nonumber
\end{align}
The operator $K^{(ij)}_{j-1\,j}$ is the exponentiation of the same two-body term in the Hamiltonian as $K_{ij}$, but acts instead on the sites $j-1$ and $j$. 

To make use of the identity above for the purpose of applying the single gate $K_{ij}$, 
one first applies the swap gates composing $\Theta_{j-1\,i}$ to the original MPS $\ket{\psi}$, yielding an effective MPS $\ket{\psi^\prime}$.
Then, the gate $K^{(ij)}_{j-1\,j}$ may be acted on $\ket{\psi^\prime}$ as a local operator. 
Finally, the swap gates composing $\Theta_{i\,j-1}$ are applied, restoring the original order of the sites. . 

Finally, when applying many Trotter gates $K_{ij}$ to an MPS, it is useful to observe that because $S(ij)^2 = 1$, the ordering of the Trotter gates in the 
decomposition can be chosen such that many of the swap gates cancel. Whenever this is the case, the repeated pair of gates can simply be omitted 
from the algorithm, leading to a significant speedup. 

One way to arrange the Trotter decomposition to take advantage of this cancellation is as follows: taking $j>i$, order the $K_{ij}$ first by $i$, from left to right 
and then for each $i$, by $j$ from left to right. Then all of the gates are applied in a single sweep that is arranged as a nested loop, where $i$ is iterated over 
in the outer loop and $j$ in the inner loop. In practice, all of the gates can be produced before applying any of them to the MPS so that cancellations can
be found and the corresponding pairs of gates omitted. Even when taking advantage of cancellations, however, this
method scales as \mbox{$L_x L_y^2 m^3$} for an \mbox{$L_x \times L_y$} ladder instead of scaling proportionally to the number of sites as in the case of a chain. 

\begin{figure}[htp]
\includegraphics[width=\columnwidth]{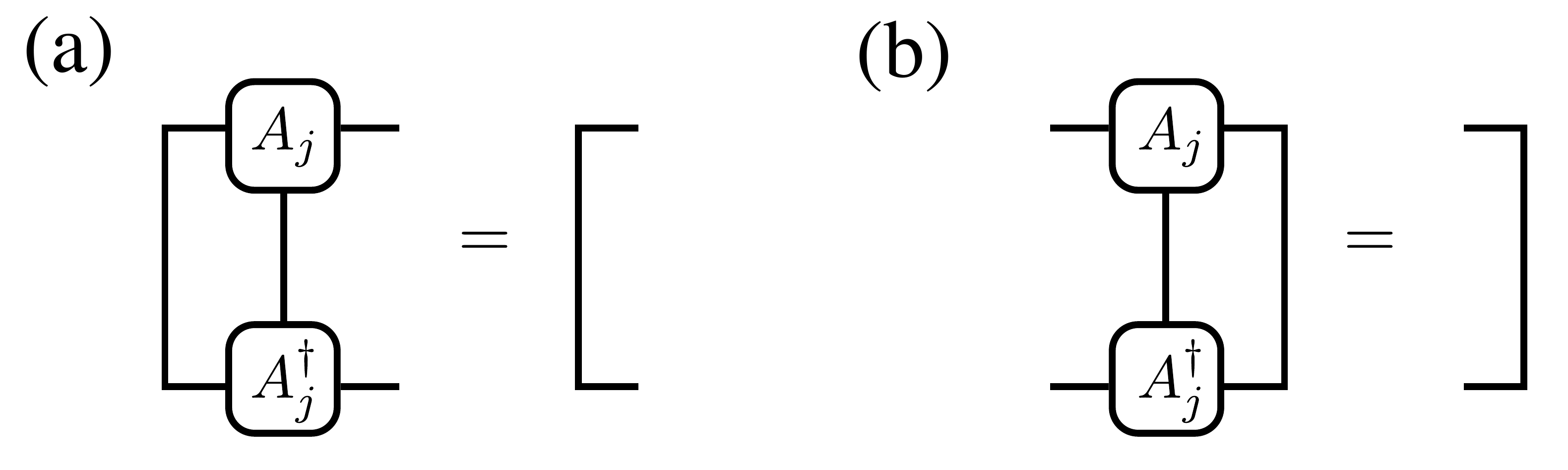}
\caption{Diagrams illustrating (a) the left orthogonality condition Eq.~(\ref{eqn:lortho}) and (b) the 
right orthogonality condition Eq.~(\ref{eqn:rortho}).}
\label{fig:orthogonality}
\end{figure}

\subsubsection{Imaginary Time Evolution with MPOs \label{sec:TimeEvolWithMPOs}}

The remaining methods for time evolution we will discuss are based upon matrix product operators, or MPOs. But, before we turn to the formation of the 
time evolution MPO itself, let us first consider how to apply it to an MPS.

A concept that will be useful is that of the orthogonality center of an MPS. If we take the center to be at site $i$, then
all of the MPS matrices $A^{s_j}$ such that $j<i$ have been made left orthogonal in the sense that 
\begin{equation}
\sum_{\alpha_{j-1}s_j} (A^{s_j})^{\dagger}_{\alpha_j^\prime\alpha_{j-1}} A^{s_j}_{\alpha_{j-1}\alpha_j} = \delta_{\alpha_j^\prime\alpha_j} \label{eqn:lortho}
\end{equation}
and all of the matrices $A^{s_j}$ such that $j>i$ have been made right orthogonal such that
\begin{equation}
\sum_{s_j\alpha_{j}} A^{s_j}_{\alpha_{j-1}\alpha_{j}} (A^{s_j})^\dagger_{\alpha_{j}\alpha_{j-1}^\prime} = \delta_{\alpha_{j-1}\alpha_{j-1}^\prime} \ . \label{eqn:rortho}
\end{equation}
These conditions are illustrated in Fig.~\ref{fig:orthogonality}. 

When an MPS has a well defined orthogonality center, the set of $A$ matrices to its left and right define an orthonormal basis for each half of the system. 
The matrix at the orthogonality center then defines the coefficients of the wavefunction in this basis, that is
\begin{equation}
\ket{\psi} = \sum_{s_i \alpha_{i-1} \alpha_i} A^{s_i}_{\alpha_{i-1} \alpha_i} \ket{\alpha^L_{i-1}} \ket{s_i} \ket{\alpha^R_i}
\end{equation}
where the site index $s_i=1,2\ldots, d$ while the bond indices \mbox{$\alpha_j=1,2\ldots, m$}. 
Alternatively, when performing two-site operations one can can take the orthogonality center to be a bond that can optionally include the sites adjacent 
to it. 

When performing the usual SVD or density matrix diagonalization in DMRG, one takes the tensor \mbox{$\psi(\alpha_{i-1} s_i s_{i+1} \alpha_{i+1})$} 
representing the wavefunction and splits it into matrices $A^{s_i}$ and $A^{s_{i+1}}$. These matrices give an optimal represention of the 
tensor $\psi$ and because they define expansion coefficients of the wavefunction in an orthonormal basis, the truncation of the wavefunction is 
also optimal. If there was no orthogonality center, or an SVD was performed away from the center, such an operation would optimally represent a local
tensor but not an optimal truncation of the wavefunction. 

In applying an MPO $W$ to an MPS $\ket{\psi_A}$ one destroys its orthogonality. 
Apart from issues of efficiency, the simplest approach to forming the product $W\ket{\psi_A}$ would be as follows. Define the matrices $B$ as
\begin{equation}
B^{s_i}_{\beta_{i-1}\beta_i} = B^{s_i}_{(\omega_{i-1}\alpha_{i-1}) (\omega_i \alpha_i)} = \sum_{s^\prime_i} W^{s_i s^\prime_i}_{\omega_{i-1} \omega_i} A^{s^\prime_i}_{\alpha_{i-1}\alpha_i}
\end{equation}
for each site $i$. If the bond dimension of $W$ is $k$ and $\ket{\psi_A}$ is $m$, this defines a new MPS $\ket{\psi_B} = W\ket{\psi_A}$ of bond dimension $mk$ 
if we think of the $\beta_i = (\omega_i\alpha_i)$ as combined, or fat, indices. 

\begin{figure}[htp]
\includegraphics[width=\columnwidth]{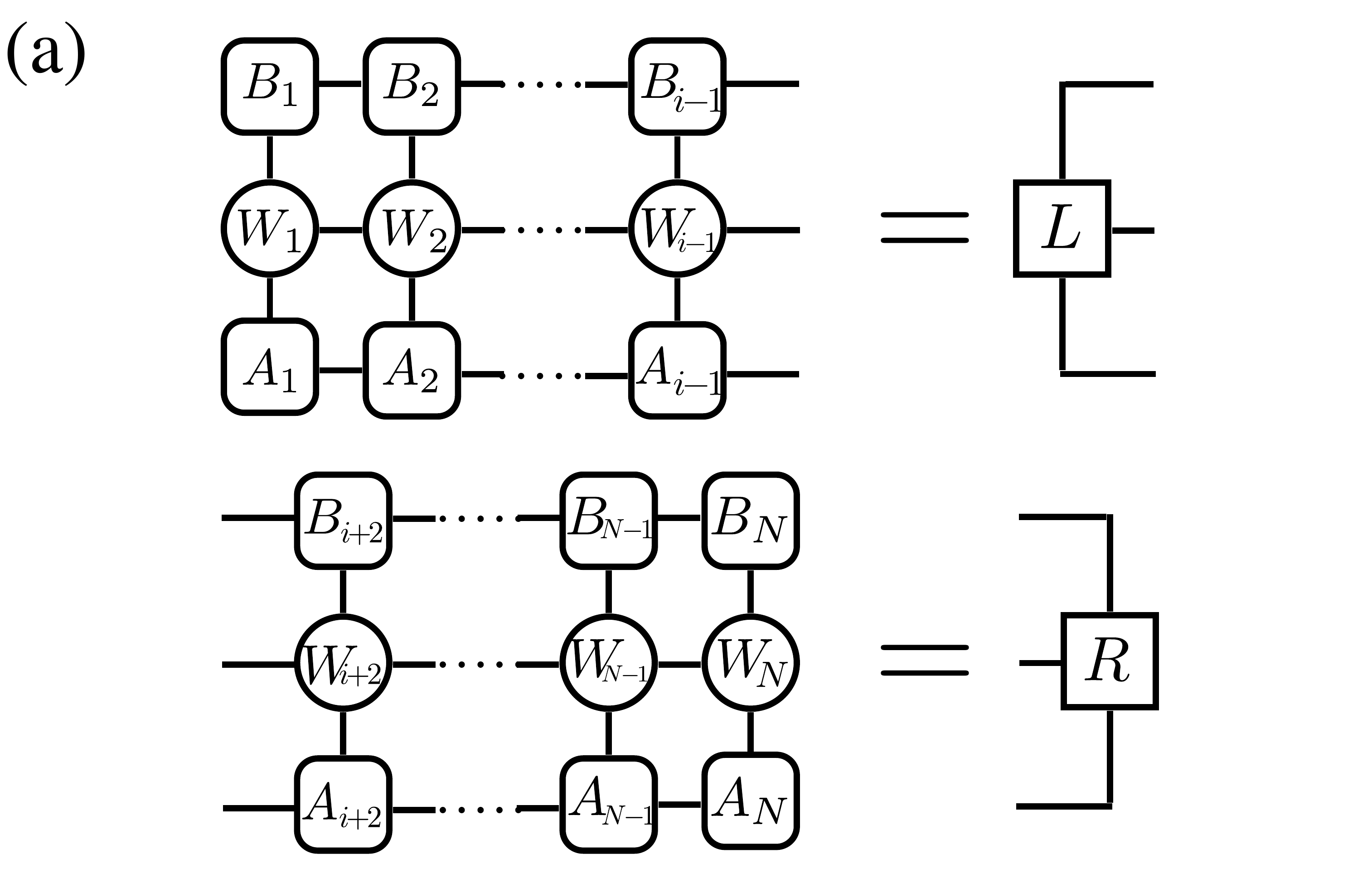}
\includegraphics[width=\columnwidth]{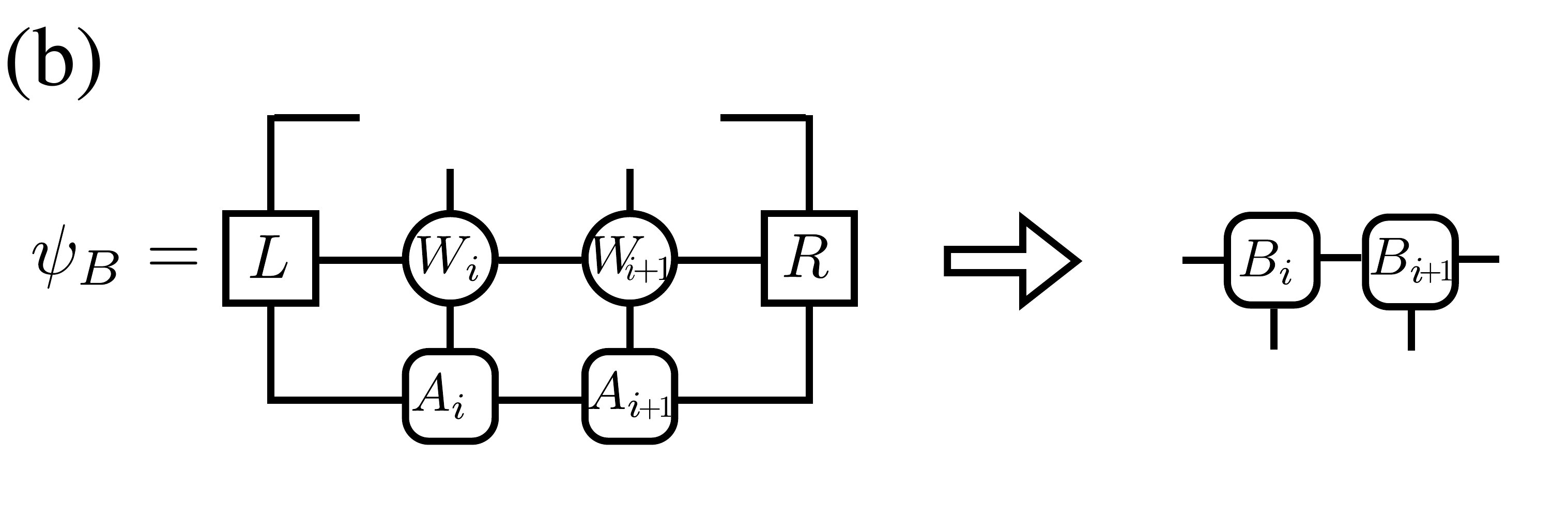}
\caption{The fitting algorithm for multiplying an MPO times an MPS. First, the tensors (a) $L$ and $R$ are formed by combining $\ket{\psi_A}$, the MPO $W$ acting on it and the resulting state $\ket{\psi_B}$ moving from the edges of the system toward the orthogonality center. Then, $\psi_B$ is defined by the tensor in (b) and split via SVD into matrices $B_i$ and $B_{i+1}$. Finally, the orthogonality center of $\ket{\psi_B}$ is shifted and the process is repeated until convergence.}
\label{fig:fitting}
\end{figure}

To guarantee proper truncation in this naive approach, one would first sweep left to right, performing SVDs to make the basis to the left orthogonal, but leaving
the bond dimension as $mk$ since the basis to the right is not orthogonal. Once at the right edge, the SVDs in the reverse sweep could truncate 
the bond dimension to $m$ or to some specified truncation error. However, this procedure scales as $N m^3 k^3$ and would be highly inefficient if $k \sim 10-100$.

Verstraete and Cirac
recommended a better approach (in a different context) that scales as $Nm^3k + Nm^2k^2$ based on fitting an MPS $\ket{\psi_B}$ to the product $W\ket{\psi_A}$.\cite{verstraete-04}
Form a random MPS $\ket{\psi_B}$ of bond dimension $m$ and orthogonalize it to have any arbitrary orthogonality center. 
The optimal two-site wavefunction $\psi(\beta_{i-1} s_i s_{i+1} \beta_{i+1})$ at this center is then found by 
forming tensors $L$ and $R$ representing the product in the basis for the left and right halves of the system and combining them with the local $W$ and 
$A$ matrices as in Fig.~\ref{fig:fitting}. This $\psi_B$ minimizes \mbox{$||\,\ket{\psi_B} - W\ket{\psi_A}||^2$}. 
One may then split $\psi_B$ into matrices $B^{s_i}$ and $B^{s_{i+1}}$  
and sweep back and forth through the system, repeating this process until it converges. 

\begin{figure}[htp]
\includegraphics[width=\columnwidth]{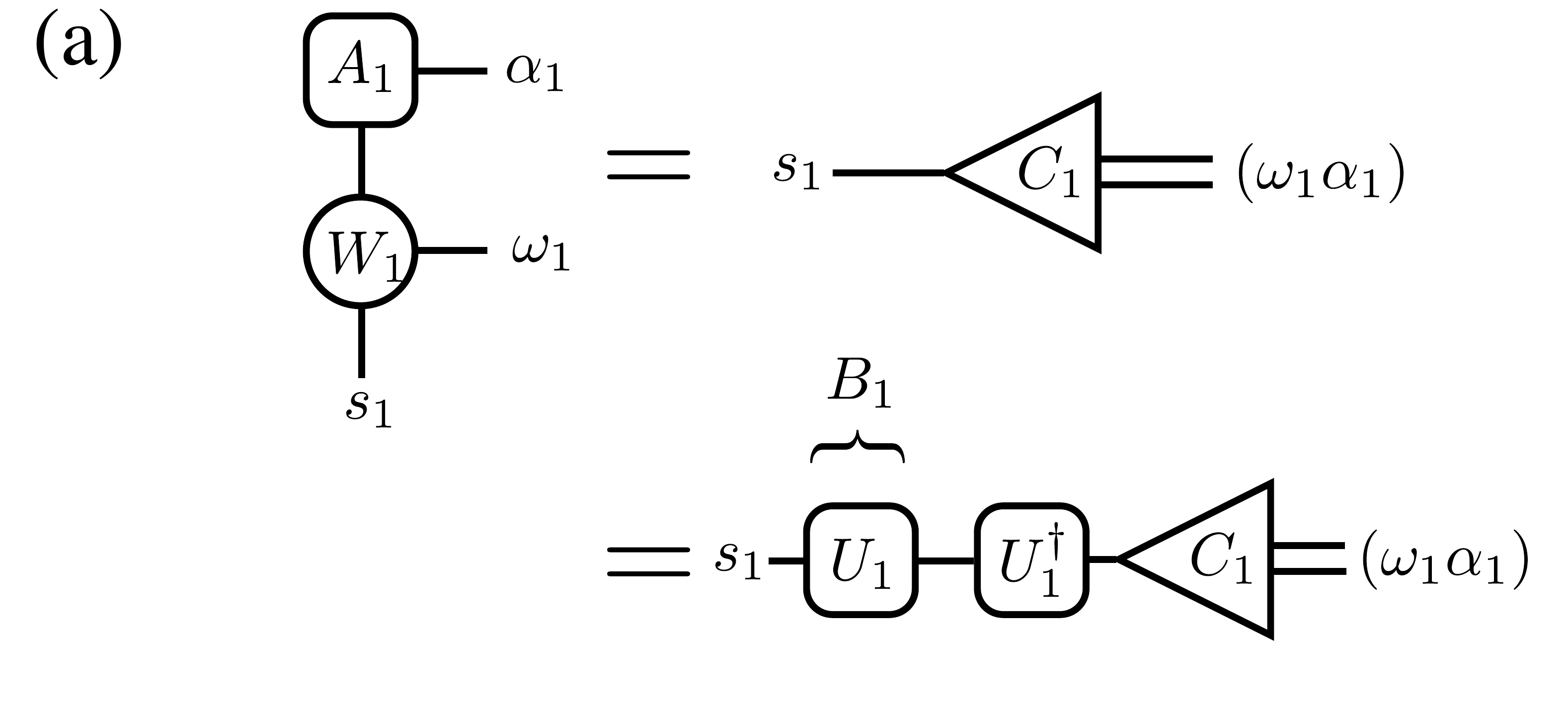}
\includegraphics[width=\columnwidth]{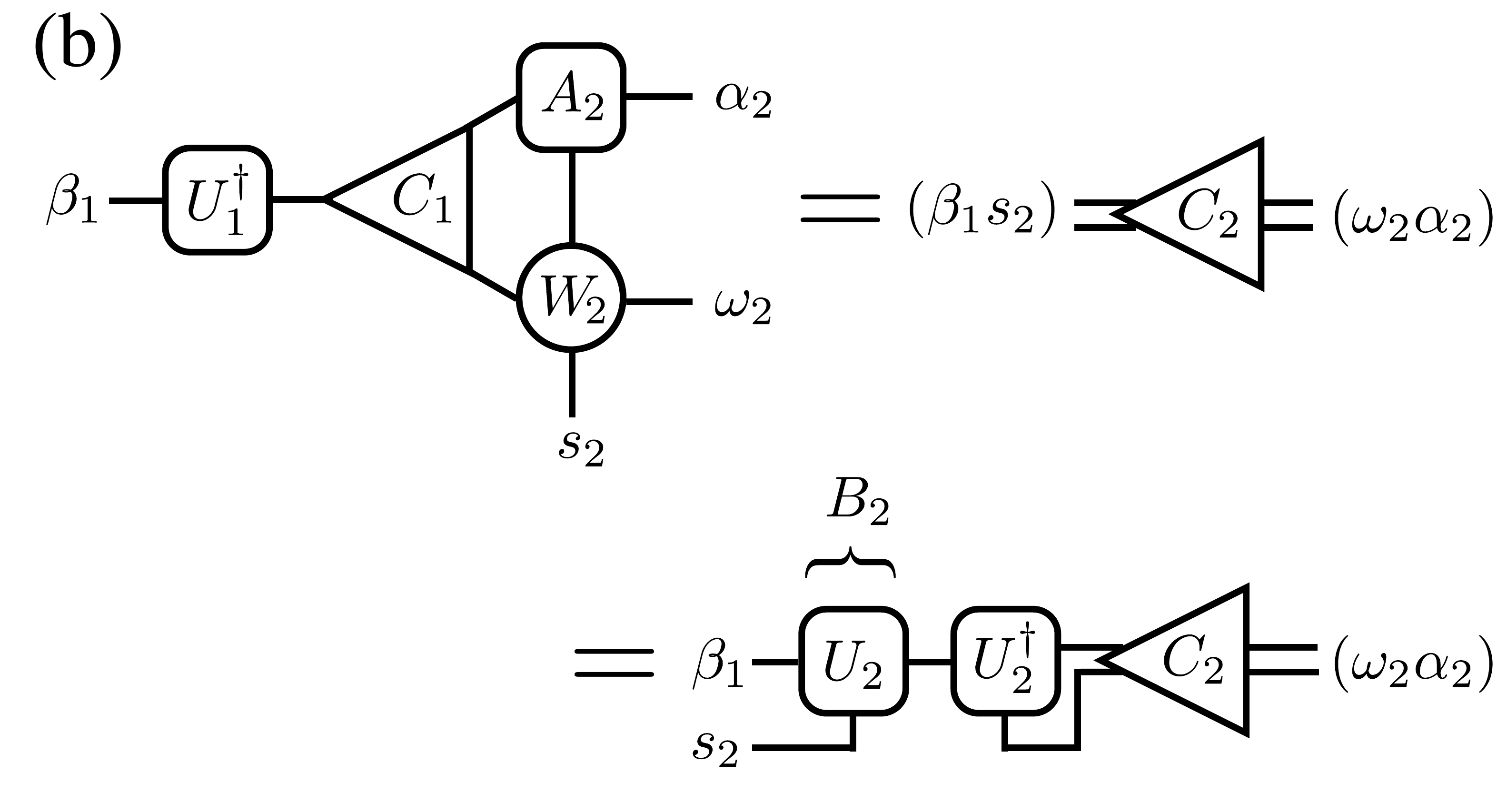}
\caption{The tensors (a) $C_1$ and (b) $C_2$ generated in the first two steps of the zip-up method for multiplying an MPO times an MPS. Indices grouped by
parentheses are to be thought of as a single fat index for the purpose of computing a Singular Value Decomposition.}
\label{fig:MPOtimesMPS}
\end{figure}

Verstraete and Cirac originally advocated a single-site version of this approach but we have found the two-site version better --- for example, 
the value of $m$ can be automatically set to achieve a specified truncation error much more readily, particularly if one is keeping track of 
conserved quantum numbers. However, even when using the two-site version, this fitting procedure may fail to converge for systems with non- nearest-neighbor interactions in a reasonable number of iterations. The reason seems to be related to the ``sticking'' problem 
of DMRG. When two distant sites are strongly entangled, the method has trouble building up this entanglement when improving the state
near each site one at a time. On the other hand, the naive method above does not have such a sticking problem. 

We have therefore developed an alternative approach, the zip-up algorithm, that has similar scaling to the fitting method but which does not get stuck. 
We begin by identifying the MPO $W$ with an MPS $\ket{W}$ by combining its primed and unprimed (i.e.\ bra and ket) site indices. 
The orthogonality centers of $\ket{\psi_A}$ and 
$\ket{W}$ are then moved to the first site. This arrangement guarantees that the product of $W$ and $\ket{\psi}$ at the first site has a basis to the right 
that, while not orthogonal, is not drastically ill-conditioned. By this we mean that while some linear dependence of the basis can exist, no basis state is 
expected to have a norm that is significantly bigger than one (norms much less than one are less of a problem). 

Then, starting from the first site, we form the product MPS $\ket{\psi_B}$ as follows.
Define the tensor $C_1$ as
\begin{equation}
C^{s_1}_{\omega_1\alpha_1} = \sum_{s_1^\prime} W^{s_1s_1^\prime}_{\omega_1} A^{s_1^\prime}_{\alpha_1} \ .
\end{equation}
Thinking of $C_1$ as a matrix $C_{s_1\,(\omega_1\alpha_1)}$, where $(\omega_1\alpha_1)$ is a combined or `fat' index, one may compute its Singular Value Decomposition (SVD) as
\begin{equation}
C_{s_1\,(\omega_1\alpha_1)} = U_{s_1\beta_1} \Lambda_{\beta_1} V^{\dagger}_{\beta_1\,(\omega_1\alpha_1)} \ .
\end{equation}
At this step, it is crucial to perform a truncation, keeping only the $m$ largest singular values contained in $\Lambda_1$. This may be done by setting $m$ explicitly or by specifying a cutoff on the truncation error.
After truncating, the first matrix of $\ket{\psi_B}$ is defined as $B^{s_1}_{\beta_1} = U_{s_1\beta_1}$ and is therefore left orthogonal by construction.

The remaining $B$ matrices may now be found recursively. To go on to the next step, one computes
\begin{equation}
C^{s_2}_{\beta_1\omega_2\alpha_2} = \sum_{s_1^\prime,\omega_1,\alpha_1,s_2^\prime} U^{\dagger}_{\beta_1 s_1^\prime} C^{s_1^\prime}_{\omega_1\alpha_1} W^{s_2s_2^\prime}_{\omega_1\omega_2} A^{s_2^\prime}_{\alpha_1\alpha_2}  \ .
\end{equation}
this time thinking of the result as a matrix $C_{(\beta_1s_2)(\omega_2\alpha_2)}$ and computing its SVD to find $U_{(\beta_1s_2)\beta_2}$. The matrix $B^{s_2}$ may then be set to $B^{s_2}_{\beta_1\beta_2} = U_{(\beta_1s_2)\beta_2}$ and the process continued by defining \mbox{$C_3 = C^{t_3}_{\beta_2\omega_3\alpha_3}$} analagously in terms of $U_2$, $C_2$, $A_3$ and $W_3$.

For clarity, the construction of the $C$ tensors is depicted graphically in Fig.~\ref{fig:MPOtimesMPS}. Asymptotically, if one assumes that $k<m$, this algorithm scales as $N m^3 k d$. 

We note that when performing the SVDs in the above procedure, it is important to use a truncation error cutoff versus a fixed $m$ 
during the initial left to right sweep to compensate for the modest loss of efficiency stemming from the fact that each truncation optimizes a 
local tensor, not $\ket{\psi_B}$ itself. 
Once the computation of $\ket{\psi_B}$ reaches the right boundary, the return SVD is fully efficient since the $B$ matrices are left orthogonal
and during this return sweep $m$ can be reduced to an optimum value. For typical 2d clusters, we find that such a procedure 
leads to an intermediate $\tilde{m}$ that is about twice $m$. 

Finally, it may be useful to combine the above two approaches. One could perform an initial left to right sweep 
using the zip-up algorithm with a fixed $\tilde{m} \simeq m$ to obtain an initial guess for $\ket{\psi_B}$ and then switch to the fitting algorithm for 
the return sweep (and possibly additional sweeps) until convergence is reached. 

\subsubsection{Taylor Series Construction of the Time Evolution MPO \label{appendix:TaylorK}}

Having discussed the application of an MPO to an MPS, let us now turn to the construction of an MPO representation of the time evolution
operator \mbox{$K_\tau = e^{-\tau H}$}. The first method we will discuss relies on expanding $K_\tau$ in a Taylor series.

To construct the MPO for $K_\tau$ using this approach, one begins by constructing an MPO representation of the Hamiltonian $H$.
While it is sometimes possible to work out the MPO for $H$ explicity, such as in the case of translationally invariant 1D systems,  
more complex systems can be treated by constructing MPOs for each term in the Hamiltonian 
and summing them together (see the Appendix for further details on adding and multiplying MPOs). 

Having found an MPO representation of $H$, the Taylor series approach begins with breaking the timestep $\tau$ into small fractions \mbox{$\epsilon = \tau/n$}.
One may then approximate \mbox{$K_\epsilon = e^{-\epsilon H}$} in a Taylor series which is truncated at a high order $p$ such that the error is controlled by $\epsilon^p$.
Finally, this series may be summed by utilizing MPO multiplication and addition algorithms. In doing so, it is important to arrange the
computational steps so as not to repeat any expensive calculations such as computing powers of $H$. One such arrangement is given by the following algorithm.

Compute the MPO $W_0 = -\epsilon  H$.
Then compute the MPOs $W_j$ recursively as
\begin{equation}
W_j =  -\epsilon H (1+\frac{1}{p-j + 1} W_{j-1})
\end{equation}
where $j=1,\ldots,p$.
The timestep operator itself is then given by  \mbox{$K_\epsilon = 1 + W_{p}$}.

Finally, the MPO for $K_\tau$ may be found by computing the $n^{\mbox{\tiny th}}$ power of $K_\epsilon$ using the zip-up method 
for MPO multiplication described in the Appendix. In doing so, it is helpful to choose $n$ to be a power of two so that $K_\tau$ may be computed by by repeatedly squaring $K_\epsilon$. In this way fewer MPO multiplications are required.

\subsubsection{Trotter-Suzuki Construction of the Time Evolution MPO \label{appendix:TrotterK}}

An alternative method for constructing the time evolution MPO $K = e^{-\tau H}\:$ is to make use of the Trotter-Suzuki decomposition of Eq.~(\ref{eqn:TSbreakup}), 
where $K$ is approximated as a product of gates $K_{ij}$. 
But instead of applying the gates directly to the MPS representing the system, one combines them together to form an MPO for $K$. 

Now, explicitly writing an MPO representation for a gate $K_{ij}$ can be challenging as it is not a simple product of single site operators and will not generally act on  sites adjacent to each other in the MPS. 
However, one solution is to use the swap gate technique described above. One first works out the set of swap gates and 
local $K^{(ij)}_{j-1\,j}$ gate operators that form $K$ and removes any redundant gates. Then the MPO for $K$ may be computed by creating an identity MPO and applying the gates directly to it, taking care to truncate the MPO bond dimension after each gate is applied. 
Or, if even greater accuracy is required, one may follow the steps above to construct an MPO for \mbox{$K_\epsilon = e^{-\epsilon H}$} with \mbox{$\epsilon = \tau/2^n$} and then use the zip-up method to square it $n$ times and obtain $K$.

An important advantage of this Trotter-Suzuki method over the Taylor series approach is that errors are easier to identify and control.
In particular, we have found that the parameters controlling the error in the Taylor series method must be reduced whenever the number of lattice sites $N$ is
increased in order to maintain the same accuracy. 
This can be traced to the fact that the Hamiltonian generally possesses at least one eigenvalue that scales as $N$, while 
other eigenvalues scale as $N^p$ with $p<1$. 

In contrast, the Trotter-Suzuki approach to computing $K$ does not suffer from this difficulty since one works with gates acting only on pairs of sites, and such gates may be constructed exactly. The only sources of error that remain, then, are the finite timestep Trotter error and the truncation error in the construction of the MPO.



\subsection{Measurement of Observables \label{sec:measurement_of_observables}}

After producing a METTS, one wants to measure a number of bulk quantities, such as the energy, as well as local observables, such as the magnetization on each site. For bulk observables that consist of local terms summed over the entire lattice, it is usually convenient to use an MPO representation.
We must therefore find an efficient way to compute expectation values of such MPOs. 

On the other hand, it is more efficient to work with local observables explicitly, and
to compute their expectation values only in terms of the orthogonality center of the MPS. And, as we will see later, this second approach is particularly
important as it is the key to performing an efficient CPS collapse in the last step of the pure state method. 

\subsubsection{Computing the Expectation Value of an MPO}

Operators formed by summing over local terms, most notably the Hamiltonian, can generally be written as an MPO with relatively small bond dimension $k$. 
Since we will need to compute the expectation value of these MPOs with respect to MPS that have a relatively large bond dimension $m$, 
let us see how to do this efficiently.

Consider computing the matrix element $\bra{\psi_A}\op{W}\ket{\psi_B}$ where $\op{W}$ is an MPO of bond dimension $k$, and $\ket{\psi_A}$ and $\ket{\psi_B}$ are MPS of bond dimension $m$. Of course, this reduces to an expectation value if \mbox{$\ket{\psi_A} = \ket{\psi_B}$}. 

One starts the computation at the beginning of the chain, defining $L_1$ as
\begin{equation}
L_{\alpha_1\omega_1\beta_1} = \sum_{s_1 t_1} (A^{s_1})^{\dagger}_{\alpha_1} W^{s_1 t_1}_{\omega_1} B^{t_1}_{\beta_1} \ .
\end{equation}
Then $L_2$ can be produced from $L_1$ by computing
\begin{equation}
L_{\alpha_2\omega_2\beta_2} = \sum_{t_2 \beta_1} \left[\sum_{s_2 \omega_1} \left[ \sum_{\alpha_1} L_{\alpha_1\omega_1\beta_1} (A^{s_2})^{\dagger}_{\alpha_1\alpha_2}\right] \!\!W^{s_2 t_2}_{\omega_1\omega_2}\right] \!\! B^{t_2}_{\beta_1\beta_2} \, .
\end{equation}
Here the order of operations in performing the sum is crucial to ensure good scaling - in particular one should not sum over both the $\alpha_1$ and $\beta_1$ 
indices simultaneously. 

One then proceeds to define $L_3$ in terms of $L_2$ in a similar manner until the end of the chain is reached. 
In this way, one computes the matrix element exactly as there is no truncation step. 
If the steps of the algorithm are followed in the order given above and we assume that $k < m$, the computation scales as $N m^3 k d$. 

\subsubsection{Computing the Expectation Value of a Local Operator}

While any operator may applied to a state by representing it as an MPO, it is often more efficient to compute the action of \emph{local} operators on an MPS explicitly. The basic idea is shift the orthogonality center of the MPS (defined in section \ref{sec:TimeEvolWithMPOs} above) such that it lies within the 
set of sites on which the local operator acts. This way any MPS matrix on which the operator acts trivially is excluded from the computation.

For example, consider computing of the expectation value of a single site operator $\op{\mathcal{O}}_i$ acting on site $i$. 
If we take the orthogonality center to be the site $i$, then because
the matrices $A^{s_j}$ for all $j<i$ obey the left orthogonality condition Eq.~(\ref{eqn:lortho})
and the matrices $A^{s_j}$ for all $j>i$ obey the right orthogonality condition Eq.~(\ref{eqn:rortho}),
the expectation value of $\op{\mathcal{O}}_i$ reduces to
\begin{equation}
\bra{\psi_A}\op{\mathcal{O}}_i\ket{\psi_A} = \!\!\! \sum_{\alpha_{i-1}\alpha_i s_i^\prime s_i} \!\! (A^{s_i^\prime})^\dagger_{\alpha_{i}\alpha_{i-1}} \bra{s_i^\prime}\op{\mathcal{O}}_i\ket{s_i} A^{s_i}_{\alpha_{i-1}\alpha_i} \: . \label{eqn:one_site_op}
\end{equation}

Or, if we are interested in applying the operator to the state to obtain $\op{\mathcal{O}}_i\ket{\psi_A}$, this can be done by making the replacement
\begin{equation}
A^{s_i}_{\alpha_{i-1}\alpha_i} \rightarrow \sum_{s_i^\prime} \bra{s_i}\op{\mathcal{O}}_i\ket{s_i^\prime} A^{s_i^\prime}_{\alpha_{i-1}\alpha_i}
\end{equation}
while keeping all other $A$ matrices fixed.

Expectation values of multi-site local operators such as $\bra{\psi_A}\op{\mathcal{O}}_i\op{\mathcal{O}}_{i+1}\ket{\psi_A}$ may be calculated in a similar way. Assuming that the 
orthogonality center is at site $i$ or $i+1$, one now includes both the matrices $A^{s_i}$ and $A^{s_{i+1}}$. However, in contrast to the single-site
computation, it is important in such multi-site expectation values to break up the tensor contractions such that the scaling of any one operation never
exceeds $m^3 d$.  

Finally, while each local expectation value may be computed with a scaling no worse than $m^3 d$, one usually wants to perform such a calculation at each site in the system. Since this necessarily involved shifting of the orthogonality center, the overall scaling will then be $N m^3 d^2$ as an SVD must be computed at each step.

\subsubsection{Estimation of Errors}

While the methods discussed above give an efficient way to calculate observables for a single METTS, one must also consider the properties of the 
METTS ensemble when estimating averages and computing error bars. This is because correlations exist both between 
sequential METTS and between related observables within a METTS.

As we will discuss in the next section, the autocorrelation time of the METTS pure state method can be made 
quite short (less than 5 steps or so) by properly choosing the basis into which the CPS are collapsed. However,
when estimating the error in a series of expectation values, it is still best to remove the effects of correlations by using a binning 
procedure as follows. 

After calculating expectation values of an observable $A$, one collects sequential values into bins whose size is larger than the autocorrelation time. 
The average value within each bin is then computed. 
Each bin average may now be taken to be statistically independent and the error estimated as the standard error of these bin averages. 

However, for quantities such as the specific heat or magnetic susceptibility, given by \mbox{$C_v = \frac{\beta^2}{N} [\avg{H^2} - \avg{H}^2]$} 
and \mbox{$\chi = \frac{\beta}{N} [\avg{m^2} - \avg{m}^2]$}, 
computing the error as the sum of the errors in the first and second moments greatly overestimates the true error even when using binning.
This can be traced to the fact that, from one step of the pure state algorithm to the next, the first and second moments of a
bulk observable $A$ are strongly correlated. That is, if $\avg{A}$ is relatively large for a particular METTS, so is $\avg{A^2}$. 

A simple way to deal with this difficulty is therefore to use a resampling method, such as the bootstrap method 
but with correlations taken into account. In other words, from the entire set of $A$ expectation values, 
one randomly samples a new set of the same size as the original but with replacement, such that a value may appear more than once.
One then calculates $\avg{A}$ for this resample. But then, instead of repeating this procedure independently for
$\avg{A^2}$, one resamples instead the \emph{same} subset of METTS. So, if expectation value of $A$ calculated from a particular 
METTS is included in the resample $n$ times, the corresponding expectation value of $A^2$ is included $n$ times as well.

The bootstrap method then consists of repeating the above process a large number of times, each time obtaining an estimate for
the second cumulant $\avg{A^2}-\avg{A}^2$. If the process is repeated enough times, the resulting distribution of second cumulant
estimates should be approximately Gaussian. Finally, the error in the second cumulant may be estimated as the standard deviation of this Gaussian distribution.


\subsection{Collapsing METTS into Classical Product States\label{sec:CPSmeasurement}}

After generating a METTS by imaginary time evolution and calculating observables of interest, the METTS must be collapsed into a CPS so that a new 
METTS can be generated with the correct probability. This can be done in any basis, a fact which one can exploit 
to reduce the autocorrelation time of the pure state algorithm and to accelerate the measurement of certain observables. 

It is also important, for reasons of efficiency, to carry out this collapse for each site one at a time. 
We will therefore discuss how to collapse a METTS in practice and show that the resulting algorithm scales as $N m^2 d^2$. 

\subsubsection{CPS Collapse Algorithm}

Since collapsing METTS is simplest for the case of a spin 1/2 model, let us begin by considering this case before 
discussing the general method. 

Say one wishes to collapse the spin of a pure state $\ket{\Psi}$ at site $i$ along the $\hat{n}$ 
axis. Define the states $\ket{+}$ and $\ket{-}$ at site $i$ such that \mbox{$\hat{n}\cdot\vec{S_i} \ket{\pm} = \pm\frac{1}{2} \ket{\pm}$}. Then the probability
of finding the spin at site $i$ to be in the $\ket{+}$ state is \mbox{$p_+ = \bra{\Psi} \hat{n}\cdot\vec{S_i} \ket{\Psi} + \frac{1}{2}$} and to be in the $\ket{-}$ state is 
$p_- = 1-p_+$.

After determining the new state for site $i$ using a random number generator, the actual collapse is enacted by the projection
\begin{equation}
\ket{\Psi} \rightarrow \left\{ \begin{array}{ll} 
p_+^{-1/2}\, \ket{+}\bracket{+}{\Psi} & \ \mbox{prob}\ p_+\\ 
p_-^{-1/2}\, \ket{-}\bracket{-}{\Psi} & \ \mbox{prob}\ p_- \ .
\end{array} \right.
\end{equation}
Finally, once this procedure is carried out for every site in the system, one obtains a CPS $\ket{\Psi^\prime}$ with probability $|\bracket{\Psi^\prime}{\Psi}|^2$.

Now consider the more general case of a lattice system where the local Hilbert space at each site $i$ can be expanded in an orthonormal basis 
$\ket{m}_i$ with $m = 1,...,d_i$. We again emphasize that this basis can be chosen arbitrarily for each site $i$ and for each step of the algorithm.

Define $P_i(m) = \ket{m}_i \bra{m}_i$ as the projection operator into the state $\ket{m}_i$.
Then the probability of finding site $i$ to be in the state $\ket{m}_i$ is just
\mbox{$p_m = \bra{\Psi} P_i(m) \ket{\Psi}$}. 

Having selecting a particular state $\ket{\tilde{m}}_i$ according to this probability, 
one completes the collapse by making the replacement
\begin{equation}
\ket{\Psi} \rightarrow p_{\tilde{m}}^{-1/2}\,P_i(\tilde{m}) \ket{\Psi} \ .
\end{equation}

Before discussing the detailed implementation of this method for MPS, let us conclude with some explicit examples of 
projection operators. From the above discussion, we can immediately identify the projectors for the $S=1/2$ basis states $\ket{+}$
and $\ket{-}$ as
\begin{align}
P_i(+) & = \frac{1}{2} + \hat{n}\cdot\vec{S}_i \nonumber \\
P_i(-) & = \frac{1}{2} - \hat{n}\cdot\vec{S}_i \: .
\end{align}
Or, if we consider an $S=1$ system and take $\ket{m}_i$ with $m=\bar{1},0,1$ to be the eigenstates of $\hat{n}\cdot\vec{S}$, 
the $P_i(m)$ may be written explicitly as
\begin{eqnarray}
P_i(1)  & = &   \frac{1}{2} (\hat{n}\cdot\vec{S_i}) + \frac{1}{2} (\hat{n}\cdot\vec{S_i})^2 \nonumber \\
P_i(0)  &  = &   1 - (\hat{n}\cdot\vec{S_i})^2 \nonumber \\
P_i(\bar{1}) & = &  -\frac{1}{2} (\hat{n}\cdot\vec{S_i}) + \frac{1}{2} (\hat{n}\cdot\vec{S_i})^2  \ .
\end{eqnarray}

\subsubsection{Collapsing an MPS Into a CPS}

Having defined the CPS collapse algorithm in a general setting, let us see how to apply it to a state $\ket{\psi_A}$ represented as an MPS.
First, assume that the orthogonality center of the MPS has been chosen to be the first site. In other words, assume that the matrices $A^{s_j}$ such that
$j=2,3,\ldots,N$ have been made right orthogonal as in Eq.~(\ref{eqn:rortho}).

One begins the collapse by first computing the expectation values of the projectors $P_1(m)$. 
As discussed in section \ref{sec:measurement_of_observables}, the existence of a well defined orthogonality center allows us to compute 
these expectation values in terms of the matrix $A^{s_1}$ alone. Explicitly, one computes
\begin{align}
\bra{\psi_A}P_1(m)\ket{\psi_A} = \sum_{\alpha_1 s^\prime_1 s_1} (A^{s^\prime_1})^\dagger_{\alpha_1} \bra{s^\prime_1}P_1(m)\ket{s_1} A^{s_1}_{\alpha_1} \nonumber \\
 = \sum_{\alpha_1} \Biggl[\sum_{s^\prime_1}(A^{s^\prime_1})^\dagger_{\alpha_1} \bracket{s^\prime_1}{m} \Biggr] \Biggl[\sum_{s_1} \bracket{m}{s_1} A^{s_1}_{\alpha_1}\Biggr] \: .
\end{align}

Once the new state of site $1$ is chosen to be $\ket{s_1} = \ket{m_1}$, say, with probability $p_{m_1} = \bra{\psi_A}P_1(m_1)\ket{\psi_A}$, the  
state could be collapsed by naively applying the projector $P_1(m_1)$ to it. However, while the action of a generic single-site operator
would be followed by an SVD to orthogonalize site $1$ and turn site $2$ into the center site, the fact that the CPS collapse operators are 
projectors allows for an even more efficient algorithm.

\begin{figure}[htp]
\includegraphics[width=\columnwidth]{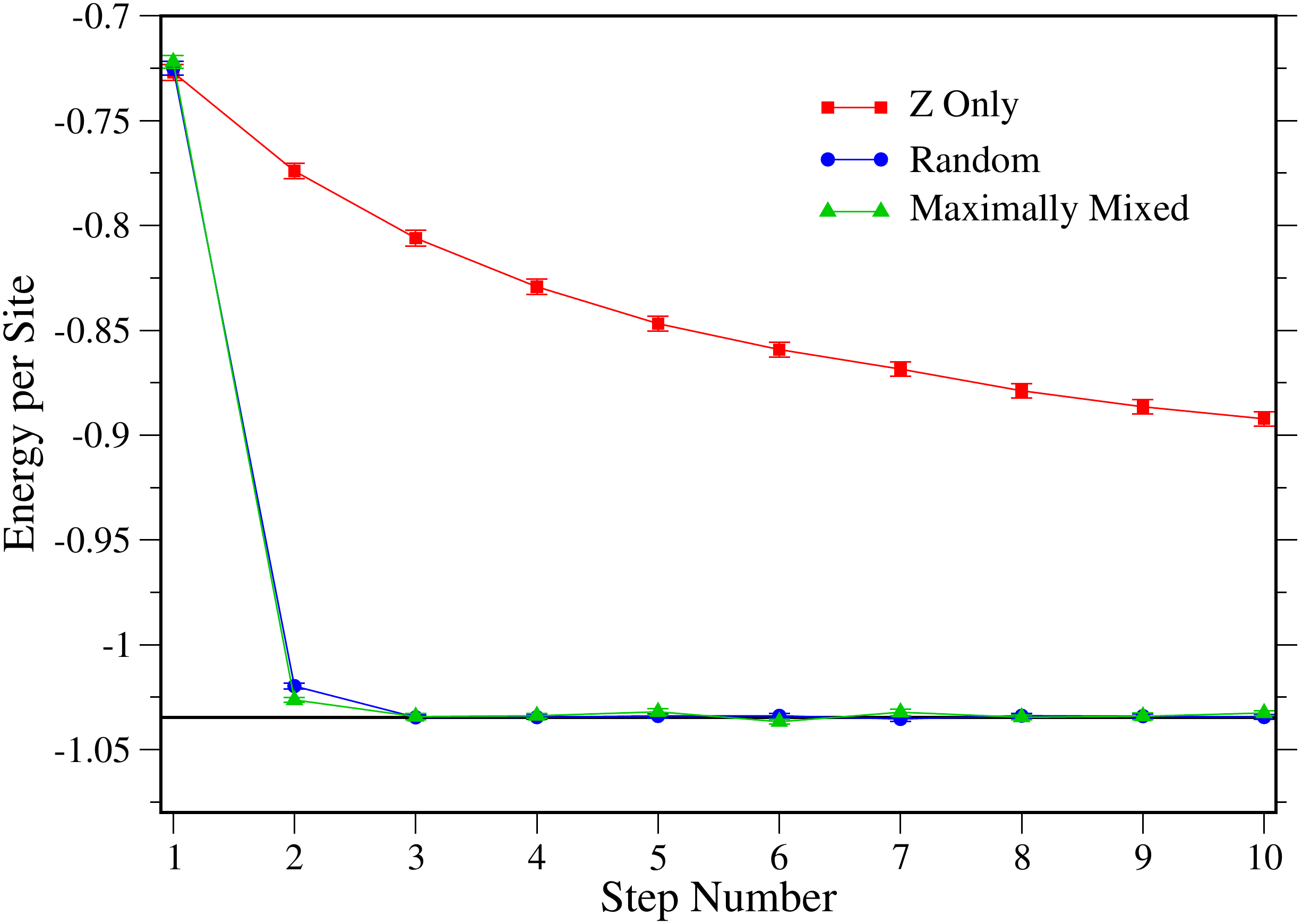} \\
\caption{The energy of the $L=50$, $S=1$ Heisenberg chain at $\beta=1.0$ found by choosing a completely random CPS (like those produced in a $\beta = 0$ simulation) and measuring the 
energy of a METTS produced after a given number of steps of the pure state method. The symbols show the ensemble average of many such runs 
each starting from a different CPS (the line is the QMC result treated here as exact). 
Depending upon the basis chosen for the CPS collapse, the number of steps needed for the effects of the initial CPS choice to 
be removed varies widely. Measuring only along the $z$ axis produces the strongest autocorrelation effects while choosing random axes or mutually
mixed bases gives much better performance and an autocorrelation time of less than 4-5 steps.}
\label{fig:energy_autocorr}
\end{figure}

Because the repeated action of single-site projectors on any state produces a product state and a product state can be represented as an MPS with 
bond dimension 1, if the above step of acting a projector $P_1(m_1)$ on site $1$ and left orthogonalizing $A^{s_1}$ were performed, the new bond index 
$\alpha_1^\prime$ connecting $A^{s_1}$ to $A^{s_2}$ could be truncated such that it only takes on one value. Anticipating this result, one can therefore 
directly replace $A^{s_1}$ with the $1\times1$ matrix \mbox{$A^{s_1} = \bracket{s_1}{m_1}$}. Finally then, to ensure that $P_1(m_1)\ket{\psi_A}$ still describes the same 
state of the remaining sites not acted on by the projector, one must replace $A^{s_2}$ according to
\begin{equation}
A^{s_2}_{\alpha_1\alpha_2} \rightarrow A^{s_2}_{m_1\alpha_2} = p_{m_1}^{-1/2} \sum_{s_1\alpha_1} \bracket{m_1}{s_1} A^{s_1}_{\alpha_1} A^{s_2}_{\alpha_1\alpha_2} \ .
\label{eqn:newA}
\end{equation}
That is, the label $m_1$ plays the role of a new bond index that runs over just one value. In fact, because it plays a trivial role, this label can be dropped
with the result that the new state of the system is a product of a single-site wavefunction for site $1$ and an $N-1$ site MPS for the remaining sites. Having made
this identification, the second site now lies on the boundary of an MPS and can be collapsed in exactly the same way as the first.

Although applying a local operator typically scales as $m^3 d^2$ per site, the considerations above show that
the CPS collapse step of the pure state algorithm can be optimized even further. The result is an algorithm that scales as $m^2 d^2$  where 
the costliest step is the formation of the new $A$ matrix in Eqn.~(\ref{eqn:newA}).

\subsubsection{Choosing the CPS Basis for Spin Models}

When collapsing a METTS into a CPS, one has the freedom to choose the CPS basis arbitrarily for each METTS and at each site.
We would therefore like to exploit this freedom first of all to ensure ergodicity and also to minimize correlations among the METTS. 

One simple strategy for improving ergodicity is to pick a 
random quantization axis $\hat{n}$ for each site at every step, and collapse the spins into the basis of $\hat{n}\cdot\vec{S}$ eigenstates.
Then if the Hamiltonian conserves a quantity such as the total $\hat{z}$ magnetization $S^z_{\text{tot}}$, such random measurements will 
generate CPS which explore many different $S^{z}_{\text{tot}}$ sectors. Moreover, random projections do not explicitly 
break rotational symmetry in the sense of favoring a particular axis from the outset. And, we have found that in practice random projections  
work quite well and give a short autocorrelation time as shown in Fig.~\ref{fig:energy_autocorr}.

In contrast, one could instead collapse every site along the same axis, say $\hat{z}$. This has a number of advantages,
including being simpler to code and making measurements of diagonal operators easier to perform, as we will discuss below. 
One particularly nice feature of using a $\hat{z}$-only basis is that, since all of the collapsed spins lie along the same axis, 
the resulting CPS can be easily visualized and plotted.

\begin{figure}[htp]
\includegraphics[width=\columnwidth]{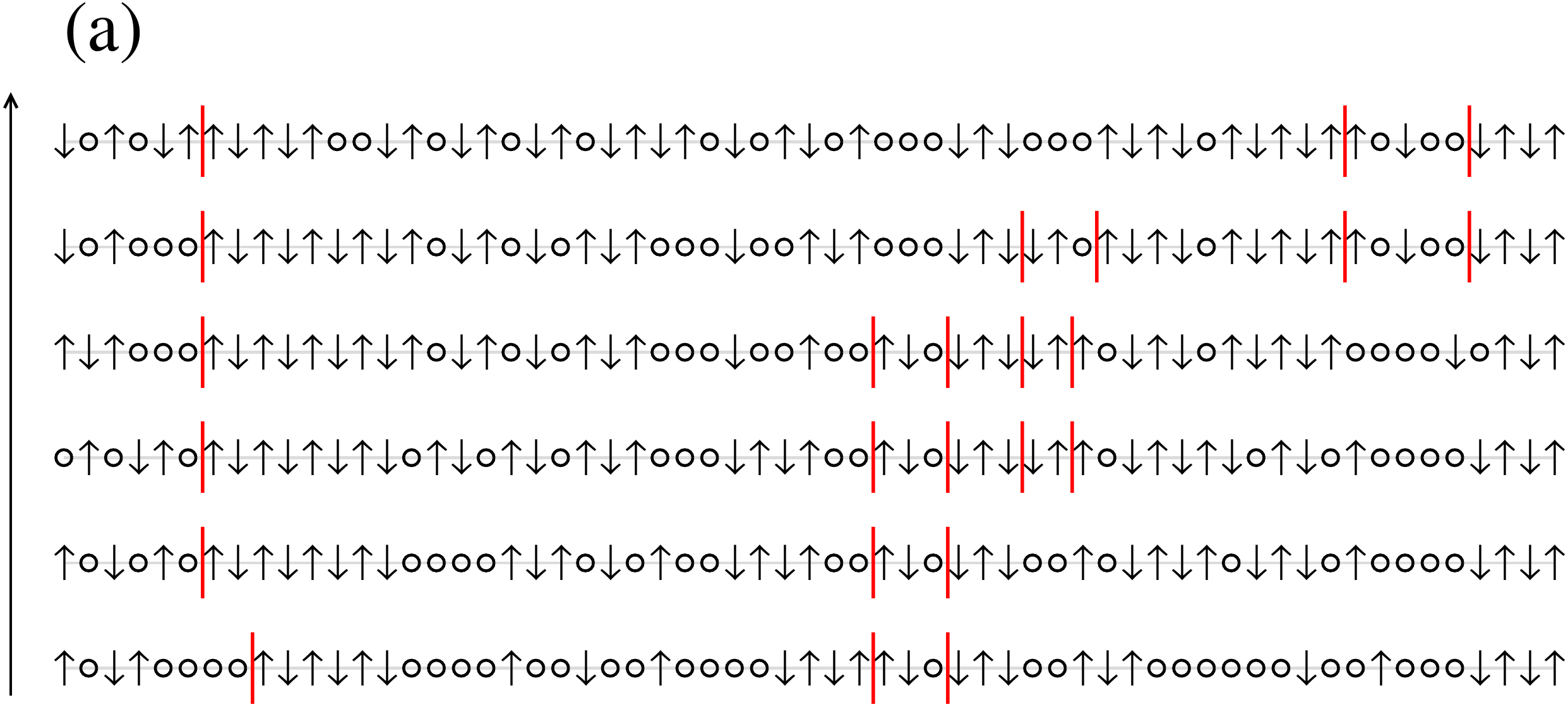} \\
\vspace{10pt}
\includegraphics[width=\columnwidth]{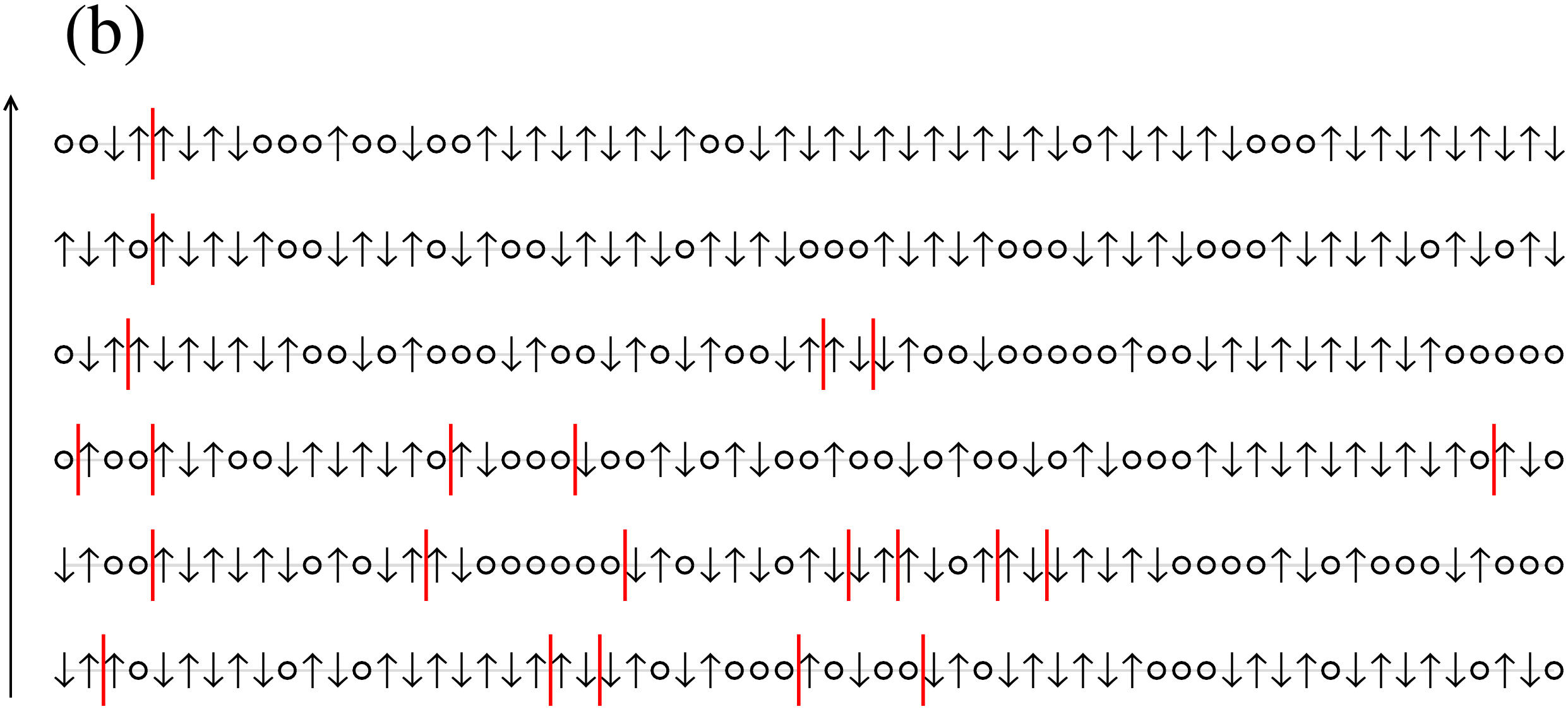} \\
\vspace{10pt}
\includegraphics[width=\columnwidth]{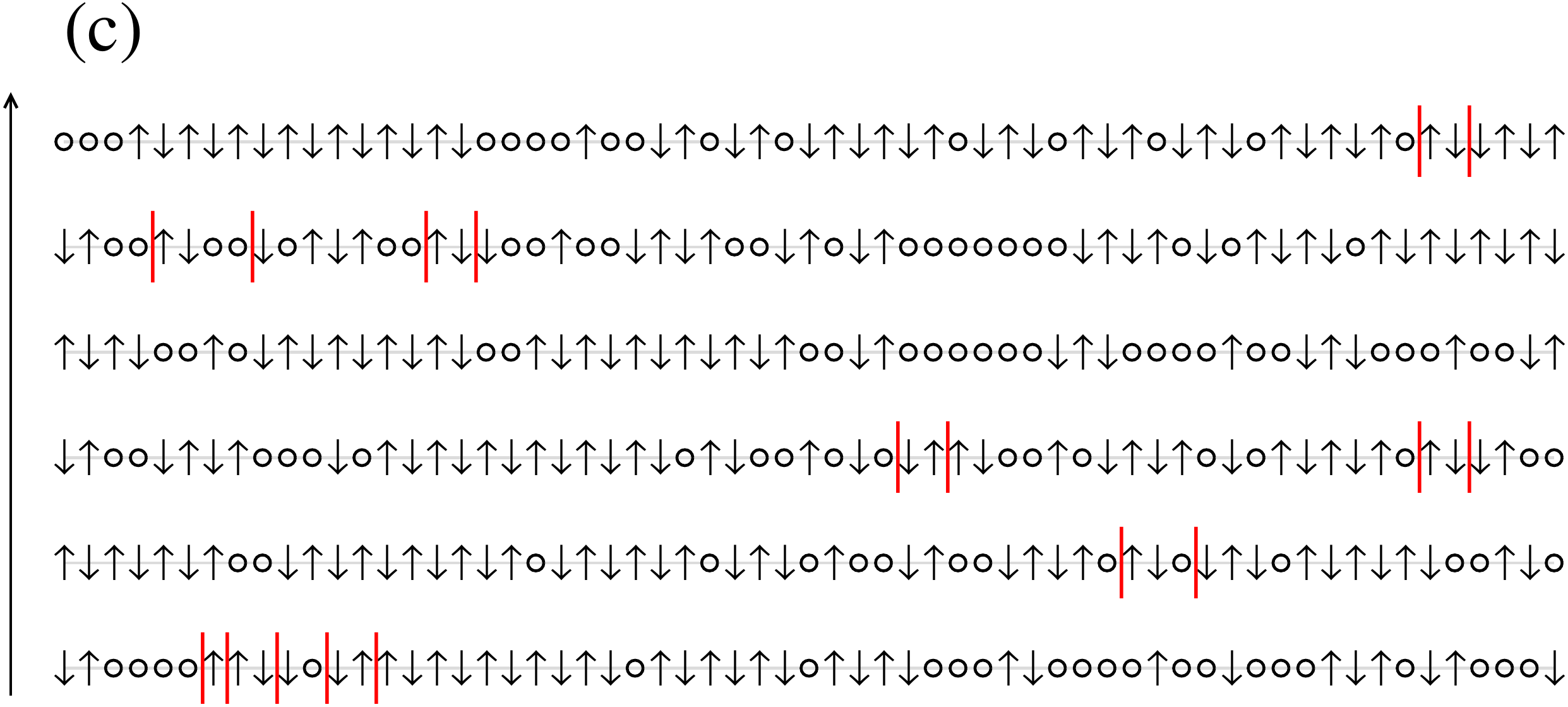}
\caption{Classical product states produced by performing \mbox{$\hat{z}$-only} measurements of METTS 
for the $S=1$ Heisenberg antiferromagnet with $N=100$. Shown are the middle 60 sites for
(a) $\beta = 1.0$, (b) $\beta = 5.0$ and (c) $\beta = 20.0$. The vertical red lines mark defects in the 
diluted N\'{e}el order of the Haldane phase.}
\label{fig:heis_seed_states}
\end{figure}

\begin{figure}[htp]
\includegraphics[width=\columnwidth]{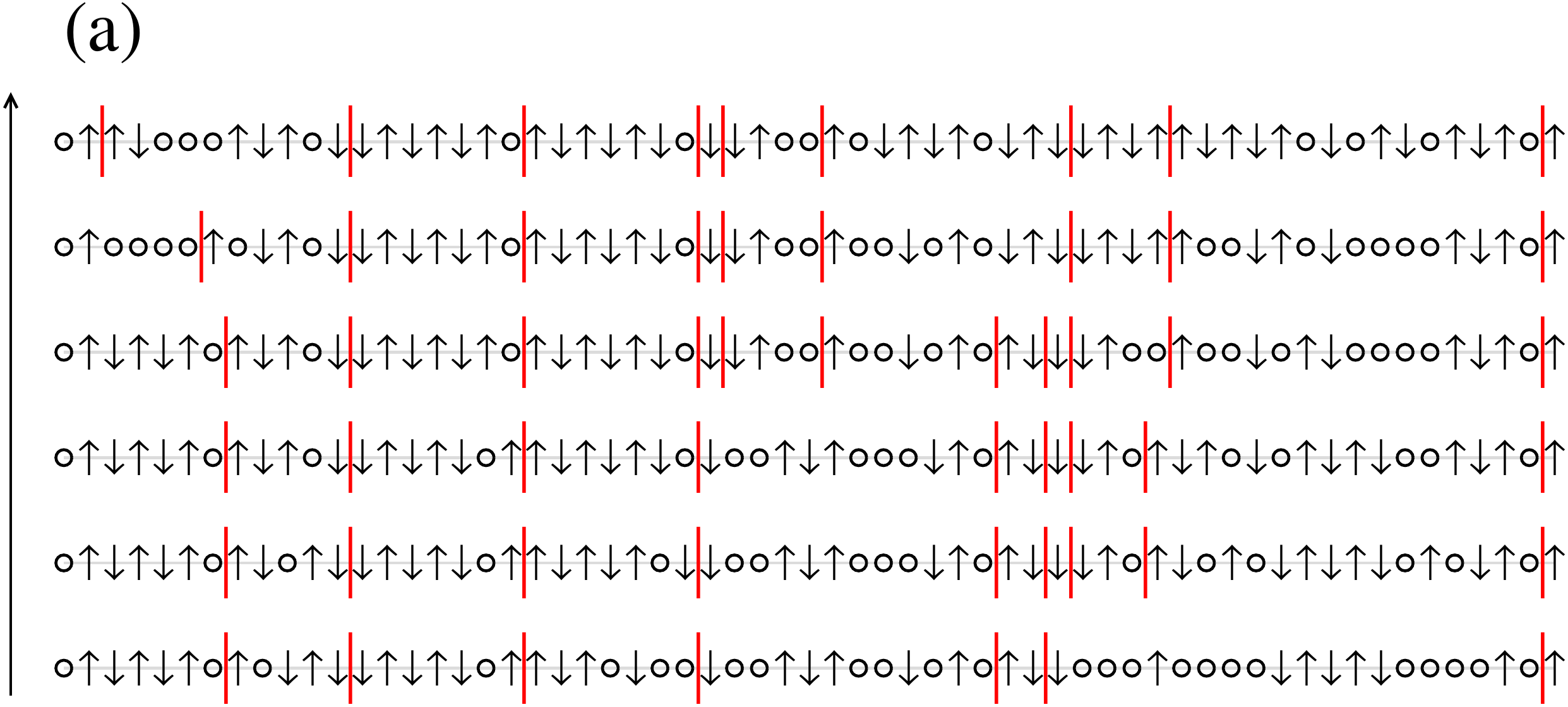} \\
\vspace{10pt}
\includegraphics[width=\columnwidth]{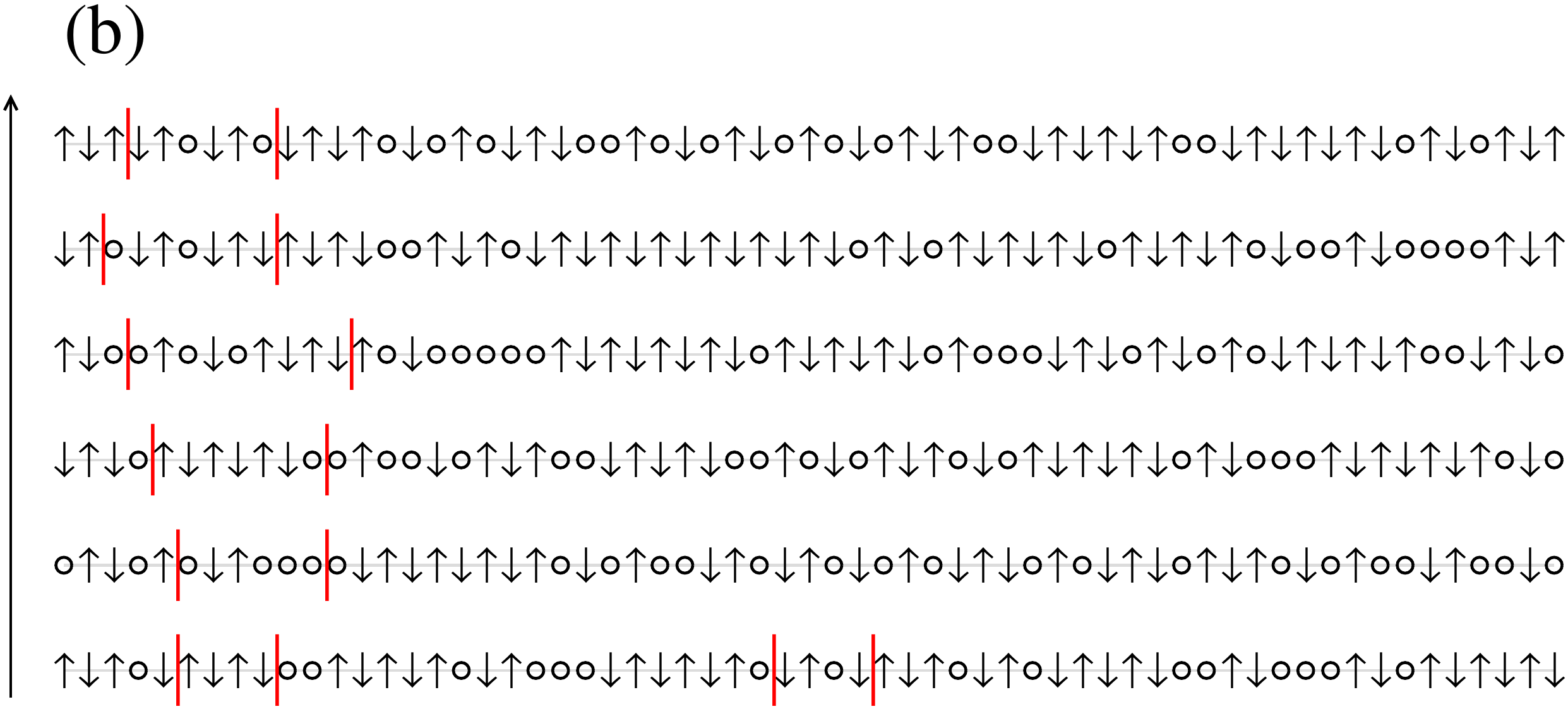} \\
\vspace{10pt}
\includegraphics[width=\columnwidth]{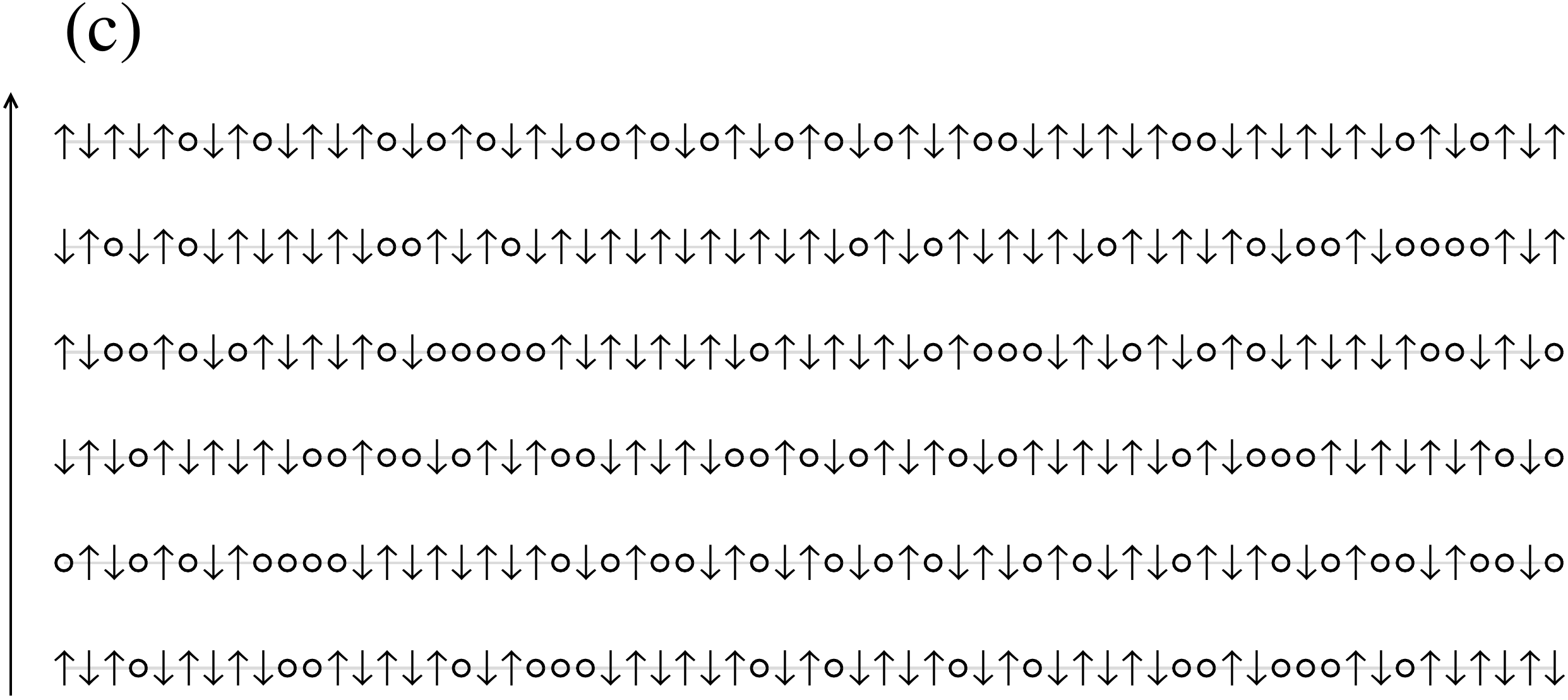}
\caption{Classical product states produced by performing \mbox{$\hat{z}$-only} measurements of METTS 
for the $S=1$ AKLT antiferromagnet with $N=100$. The temperatures and symbols are identical to those in Fig.~\ref{fig:heis_seed_states}.}
\label{fig:aklt_seed_states}
\end{figure}

For example, in Figs.~\ref{fig:heis_seed_states} and \ref{fig:aklt_seed_states} we show the CPS obtained after
collapsing subsequent METTS in simulations of the Heisenberg and AKLT models and mark points at which the spins fail to 
follow the diluted N\'{e}el pattern that underlies the
string order in the Haldane phase. It is interesting to see the number of defects decrease as the temperature is 
lowered, and to observe that for the AKLT case, they disappear entirely as $T\rightarrow 0$. 

A serious drawback to the $\hat{z}$-only approach, however, is that it leads to problems with
ergodicity. In practice, using the $\hat{z}$ basis only tends to lengthen the autocorrelation time, 
reducing the efficiency of the algorithm for calculating thermal averages. For example, one can clearly
see strong correlations in the positions of the defects plotted in Figs~\ref{fig:heis_seed_states}a
and \ref{fig:aklt_seed_states}a. And, correlations in energy measurements can persist over many steps as shown in Fig.~\ref{fig:energy_autocorr}.


Since we would like to retain the advantages of the $\hat{z}$-only basis but cannot afford to sacrifice ergodicity, one
solution is to collapse METTS along the $\hat{z}$ axis only on even numbered steps. If we then collapse along random axes after odd steps, we can 
maintain a short autocorrelation time. 

This compromise suggests an even better approach, however. Instead of switching to random axis CPS collapses, we can use a basis that is maximally
mixed relative to the $S^z$ eigenstates. For example, in a $S=1/2$ system one would perform odd step collapses along the $\hat{x}$ axis.
As $S^{x}$ eigenstates are in an equal superposition of $S^z$ eigenstates, the autocorrelation time is expected to be minimal since 
any memory of the previous CPS will be completely erased.

Likewise, for $S=1$ systems, one can perform odd step collapses in the maximally mixed basis given by
\begin{align}
\ket{\mu_1} & = \frac{1}{\sqrt{3}} (e^{i 2\pi/3}  \ket{1} - \ket{0} - e^{-i 2\pi/3}  \ket{\bar{1}}) \nonumber \\
\ket{\mu_2} & = \frac{1}{ \sqrt{3}} (\ket{1} - \ket{0} - \ket{\bar{1}}) \nonumber \\
\ket{\mu_3} & = \frac{1}{\sqrt{3}} (e^{-i 2\pi/3}  \ket{1} - \ket{0} - e^{i 2\pi/3}  \ket{\bar{1}}) \: .
\end{align}
As shown in Fig.~\ref{fig:energy_autocorr}, using this basis on odd steps reduces correlation effects even more effectively than the 
random axis approach, for which the autocorrelation time is already quite short.

\subsubsection{Diagonal Measurements with Product States}

While the cost of computing expectation values of quantities such as the energy or the magnetization is quite low for each
METTS produced, other measurements can be very costly.

For example, we may want to measure real space spin correlations \mbox{$C(\Delta) = \sum_i \avg{\vec{S}_i\cdot\vec{S}_{i+\Delta}}$}. 
One approach would be to create an MPO that encodes the sum over all pairs of spin operators separated by $\Delta$.
Or, we can work with MPOs for the individual spin operators and perform the sum above for each METTS. 
In practice however, we have found that both approaches are very costly because of both the time needed to construct the MPOs and to compute 
the expectation values.

But, assuming we are studying a spin rotationally invariant system, only the $\avg{S^z_i S^z_j}$ correlations are needed
as the others are equal by symmetry. And, since $S^z_i S^z_j$ is diagonal in the product basis of $S^z$ eigenstates, we can 
take advantage of this fact to perform correlator measurements using the CPS collapsed from each METTS instead of using the METTS
themselves. As long as the operator we wish to calculate is diagonal in the basis of the collapsed CPS, we will still obtain the correct thermal average.

So, if one performs $\hat{z}$-only collapses as discussed in the previous section, the operator content of measuring an diagonal observable such as 
\mbox{$C^z(\Delta) = \sum_i \avg{S^z_i S^z_{i+\Delta}}$} can be ignored entirely. That is, if we collapse a METTS into a CPS $\ket{\Psi}$ in the $\hat{z}$ basis such that
\begin{equation}
\ket{\Psi} = \ket{s_1} \ket{s_2} \ket{s_3} \ldots \ket{s_N} \: ,
\end{equation}
then we immediately have $C^z(\Delta) = \sum_i s_i \: s_{i+\Delta}$.

\begin{figure}[htp]
\includegraphics[width=\columnwidth]{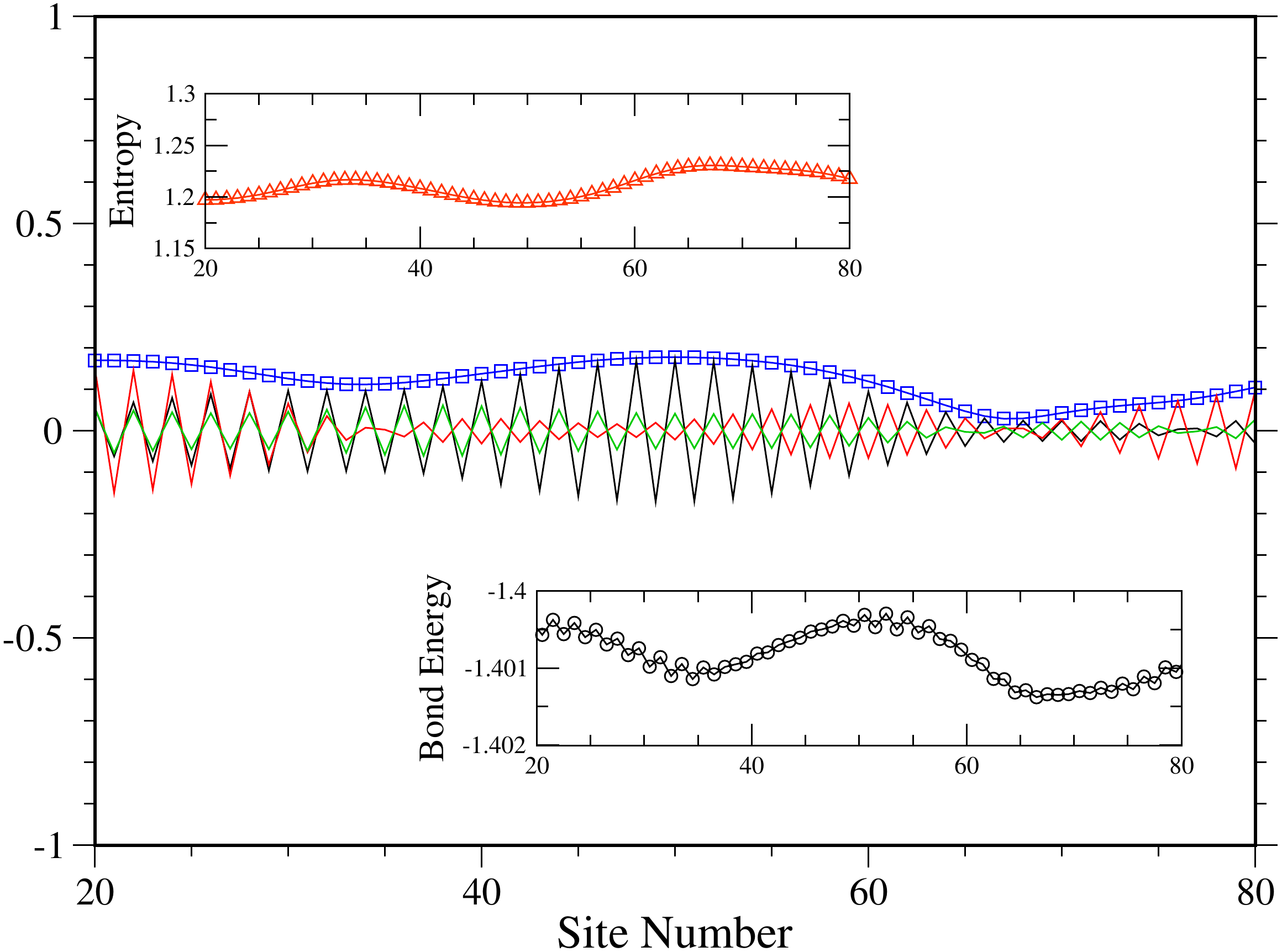}
\caption{Properties of a METTS produced for the 100 site $S=1$ Heisenberg chain at $T=0.1$, central 80 sites. 
In the main plot, the solid lines
(red, green and black) show the 3 components of $\avg{\vec{S}}$ while the (blue) boxes show $|\avg{\vec{S}}|$. 
The entanglement entropy on each bond is shown in the top inset, while the expectation value of each Hamiltonian bond term is shown
in the bottom inset.}
\label{fig:heis_typical}
\end{figure}

One can make even better use of the resources expended in producing each METTS by 
re-collapsing it multiple times and measuring each resulting CPS. This procedure also produces the correct thermal average for diagonal 
observables and 
is much more efficient for large $\beta$ than collapsing each METTS only once. By using the pure state algorithm in this way,
one explicitly samples not only the thermal fluctuations, but the quantum fluctuations as well.

Finally, since the above method relies on performing CPS collapses along the $\hat{z}$ axis, it is susceptible to the same ergodicity issues mentioned earlier. However, ergodicity can be restored by alternating between $\hat{z}$ axis
collapses on even steps and random or maximally mixed collapses on odd steps.
And, for the $S=1/2$ case, since the maximally mixed basis is just the $\hat{x}$ basis, we can perform measurements of quantities like 
$C(\Delta)$ for rotationally invariant models on $\emph{every}$ step by treating the $\hat{x}$ basis as an effective $\hat{z}^\prime$ basis.

\begin{figure}[htp]
\includegraphics[width=\columnwidth]{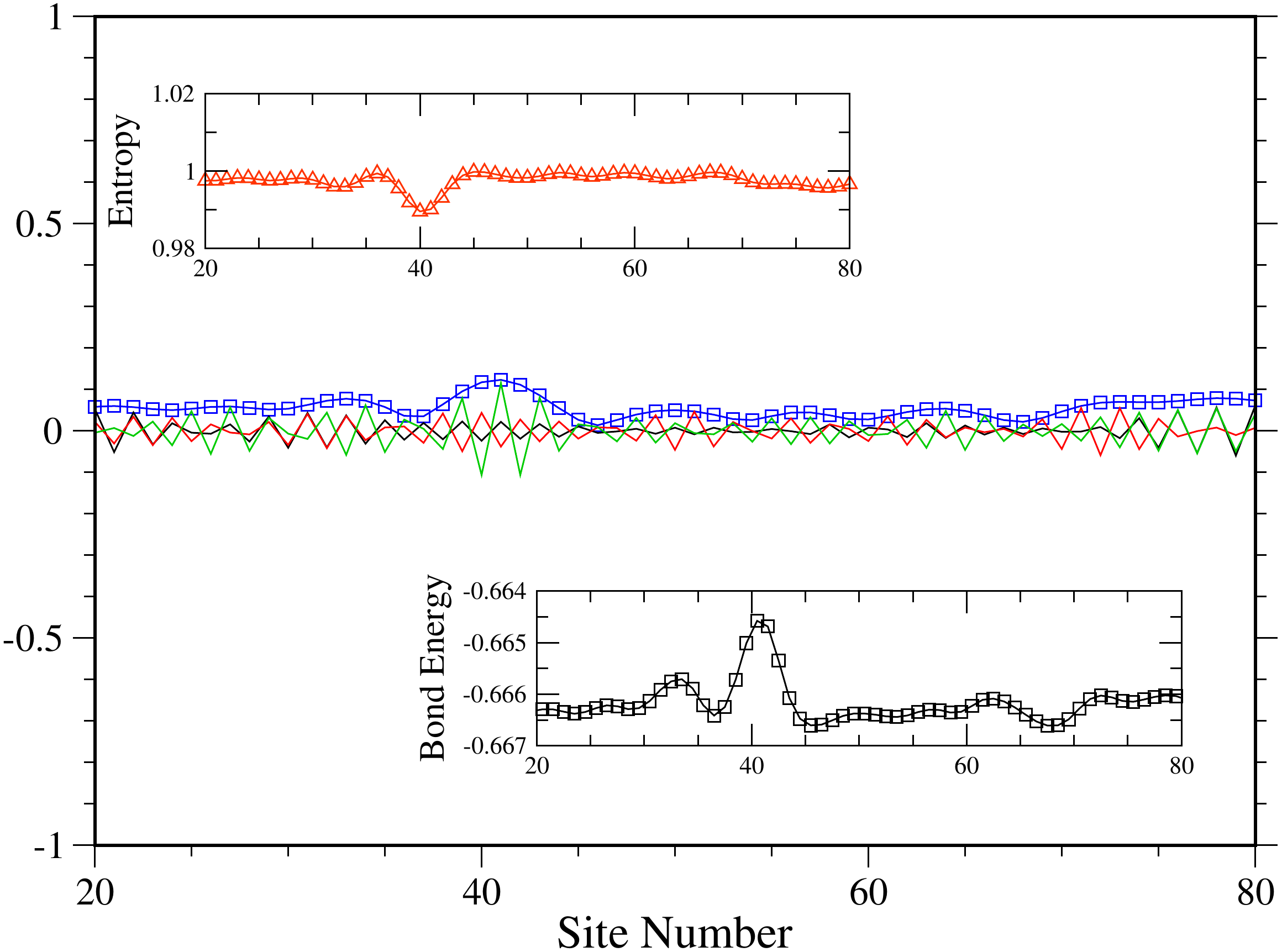}
\caption{Properties of a METTS produced for the 100 site $S=1$ AKLT chain, central 80 sites. The temperature and symbols are the same as those in
Fig.~\ref{fig:heis_typical}.}
\label{fig:aklt_typical}
\end{figure}

%
%

\section{Properties of METTS \label{sec:properties}}

While the pure state algorithm is effective for computing thermal averages, it also generates an entire wavefunction
at each step that may be thought of as a snapshot of the system. Every METTS it produces can be used to 
simultaneously calculate many different observables and look for characteristic fluctuations or short range order.

Studying the properties of METTS can also help us to understand why relatively few METTS
are needed to accurately estimate observables. By identifying the METTS ensemble which maximizes this 
sampling efficiency for a simple model, we will discover that while the entanglement entropy of such an ensemble is not
always optimal, it may be minimal in a more restricted sense.  
 
\subsection{Observing Thermal Fluctuations With METTS}

Figures \ref{fig:heis_typical} and \ref{fig:aklt_typical} each show the properties of a single METTS produced for the spin one Heisenberg and 
AKLT chains, respectively.
The temperature is taken to be \mbox{$T=0.1$} in the units of Eqn.~\ref{eqn:S1Ham}.

Throughout both chains, antiferromagnetic spin correlations are clearly visible but thermal fluctuations are also apparent.
For example, the AKLT chain in Fig.~\ref{fig:aklt_typical} shows longitudinal fluctuations in the magnitude of $\avg{\vec{S}}$ around site 40. In the inset, one can
see that the local bond energy is also correspondingly higher at this point in the chain. This may be understood as a fluctuation away from 
the spin liquid ground state in which \mbox{$\avg{\vec{S}} = 0$}.

A second type of thermal fluctuation that is visible is a twist in the staggered magnetization, such as the one near site 30 of the 
Heisenberg chain of Fig.~\ref{fig:heis_typical}. 
Near this site, the spin length is also shortened and the local bond energy is relatively low. After the twist, a second region of correlated spins persists 
for a few correlation lengths until another twist occurs near site 70.

It is interesting to observe that since each METTS is a pure state, the von Neumann entanglement entropy for bipartitions of 
the system about each bond is well defined, and we normalize it here such that a spin 1/2 singlet has an entropy of 1. As Figs.~\ref{fig:heis_typical} and \ref{fig:aklt_typical} show, 
in the regions where the spins are more classical and the bond energy is relatively higher, the entanglement entropy is lower, signifying weaker quantum
fluctuations.

\subsection{Energy Measurements and the Efficiency of METTS}
\begin{figure}[htp]
\includegraphics[width=\columnwidth]{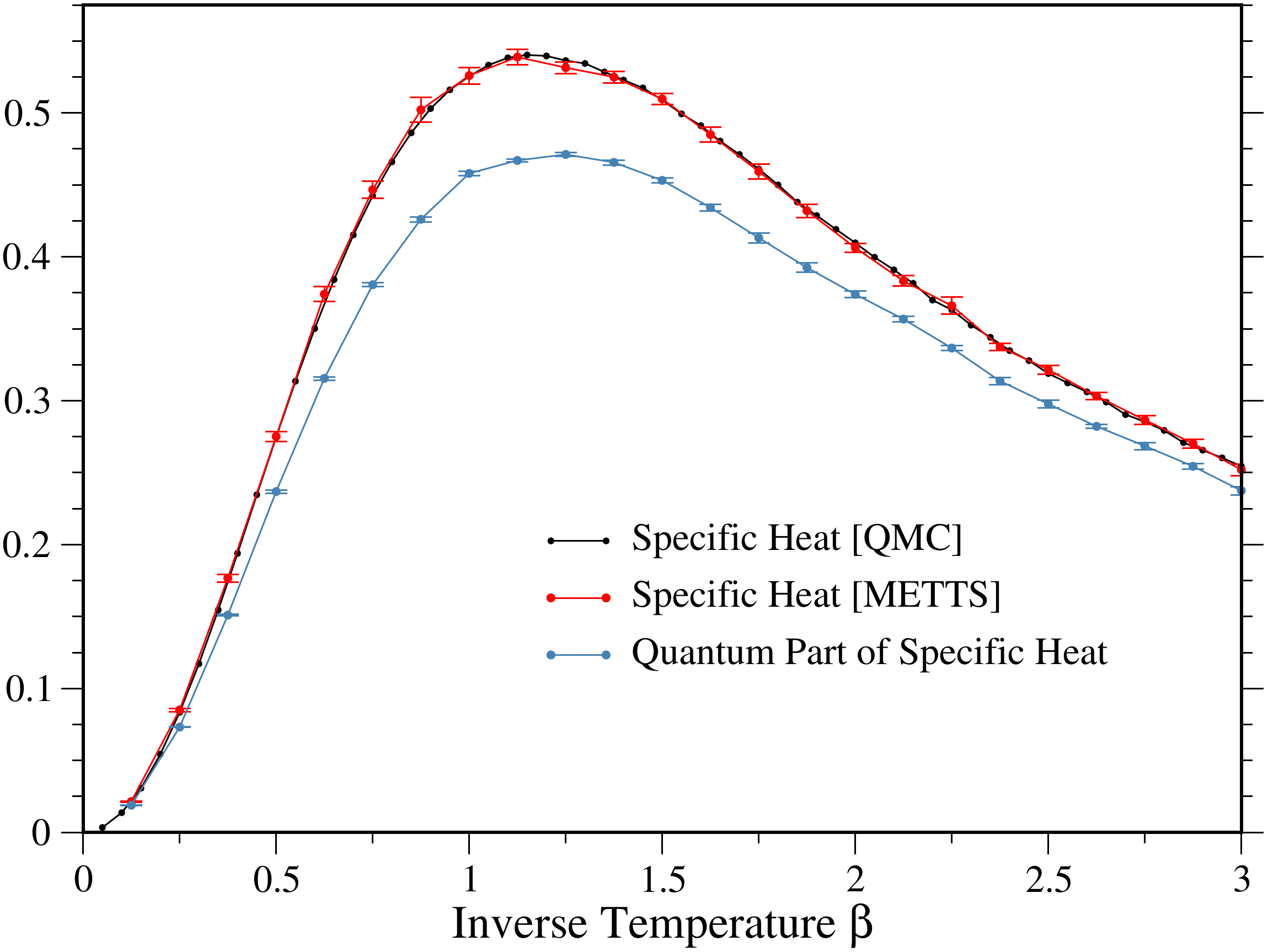}
\caption{The specific heat of the $S=1$ Heisenberg chain with $N=100$ sites calculated using METTS with 
each CPS collapse performed along random axes. Also shown is the quantum specific heat, that is, 
the average estimate of the specific heat as calculated from a single METTS.}
\label{fig:quant_spec_heat_comparison}
\end{figure}

Let us now turn our attention from the properties of individual METTS to ensembles of METTS. An important concept will be 
that of a \emph{thermal decomposition}, by which we mean a set of pure states $\ket{\phi(i)}$ and weights $P(i)$ such that
\begin{equation}
\sum_i P(i)\: \ket{\phi(i)}\bra{\phi(i)} = e^{-\beta H} \: . \label{eqn:decomposition}
\end{equation} 
In fact, all density matrices can be decomposed in this way and the $\ket{\phi(i)}$ need not be orthogonal, though they should be normalized. 
From Eqs.~(\ref{eqn:phi_state}) and (\ref{eqn:weight}), one sees that every METTS ensemble corresponds to such a decomposition.

Implicit in Eq.~(\ref{eqn:decomposition}) is a division of thermal effects into classical
fluctuations quantified by the weights $P(i)$, and quantum fluctuations present within the states $\ket{\phi(i)}$,
although this distinction is somewhat arbitrary as there exist many different thermal decompositions. 
But, because we are interested in sampling the $\ket{\phi(i)}$ rather than calculating them all, 
our sampling process will be especially efficient when most of the thermal effects can be captured within the states themselves, rather than in their distribution. 

To make this intuition more precise, let us consider the energy fluctuations within a given thermal decomposition.
Using the notation \mbox{$\avg{A}_i = \bra{\phi(i)} A \ket{\phi(i)}$}, we may write the specific heat as 
\begin{equation}
C_v = \frac{\beta^2}{N} \left[ \sum_i \frac{P(i)}{\mathcal{Z}} \avg{H^2}_i - \left(\sum_i \frac{P(i)}{\mathcal{Z}} \avg{H}_i\right)^2\: \right] \,.
\end{equation}
However, the fact that we are working with an ensemble of pure states also allows us to define a `quantum specific heat'
\begin{equation}
C^{(q)}_v = \frac{\beta^2}{N} \sum_i \frac{P(i)}{\mathcal{Z}} \biggl[\avg{H^2}_i - \avg{H}_i^2\biggr]
\end{equation}
that quantifies how much each state $\ket{\phi(i)}$ contributes to the full specific heat on average. 
 
Now it turns out that, up to a constant factor, the difference between the quantum specific heat and 
the total specific heat is nothing but the variance of the expectation value of the energy within the ensemble, that is
\begin{align}
C_v - C^{(q)}_v & = \frac{\beta^2}{N} \sigma^2(H) \\
& = \frac{\beta^2}{N} \left[ \sum_i \frac{P(i)}{\mathcal{Z}} \avg{H}_i^2 - \left(\sum_i \frac{P(i)}{\mathcal{Z}} \avg{H}_i\right)^2\: \right] \: . \nonumber
\end{align} 
So, the smaller this difference, the fewer the number of states that must be sampled to estimate the energy to a given accuracy.

Now turning to a particular choice of a thermal decomposition, let us first consider the exact energy eigenstates of the system, taking 
\mbox{$\ket{\phi(i)} = \ket{\epsilon_i}$} and \mbox{$P(i) = e^{-\beta \epsilon_i}$}. Since $H$ is diagonal in this basis, it follows that \mbox{$\avg{H^2}_i = \avg{H}_i^2$}.
Therefore the quantum specific heat is precisely zero and, as a result, the variance in the energy is maximal. So even if one 
could sample the exact energy eigenstates of the system, they would make an exceedingly poor basis for the purpose of estimating the 
average energy. 

Next, consider a decomposition of the thermal ensemble in terms of METTS. From the calculation of both the specific heat and 
the quantum specific heat shown in Fig.~\ref{fig:quant_spec_heat_comparison} for the $S=1$ Heisenberg chain, we can see that, even near the
peak where the difference between $C_v$ and $C^{(q)}_v$ is greatest, the quantum specific heat is only smaller than the total by about a factor of $0.8$.  
This indicates that the METTS decomposition we sampled is quite close to being optimal. 
Physically, we may attribute this efficiency to the fact that, for a given temperature $T$, 
each METTS has a significant overlap with all of the energy eigenstates $\ket{\epsilon_i}$ such that \mbox{$\epsilon_i \lesssim T$}. 

Finally then, let us see what would happen if we use a decomposition of the thermal ensemble in which this overlap with the energy eigenstates is
maximal. Consider the overcomplete, unnormalized basis of states
\begin{equation}
\ket{\Theta} = \sum_i e^{i\theta_i} \ket{\epsilon_i}
\end{equation}
each parameterized by the set $\Theta = \{\theta_i\}$. Then the normalized states given by
\begin{equation}
\ket{\phi(\Theta)} = \frac{1}{\sqrt{\mathcal{Z}}} e^{-\beta H/2}\ket{\Theta} =  \frac{1}{\sqrt{\mathcal{Z}}}  \sum_i e^{i\theta_i} e^{-\beta \epsilon_i/2} \ket{\epsilon_i}
\end{equation}
form a decomposition of a thermal ensemble with uniform weights $P(\Theta) = \text{const.}$
If we now imagine sampling even just one such state, the resulting estimate of the energy 
would be exact, since
\begin{align}
\bra{\phi(\Theta)} H \ket{\phi(\Theta)} & = \frac{1}{\mathcal{Z}} \sum_{i,j} e^{i(\theta_j - \theta_i)} e^{-\beta (\epsilon_j + \epsilon_i)/2} \bra{\epsilon_j} H \ket{\epsilon_i} \nonumber \\
& = \frac{1}{\mathcal{Z}} \sum_i e^{-\beta\, \epsilon_i}\: \epsilon_i \nonumber \\
& = \avg{H} \:.
\end{align}
In a similar manner, one may obtain the exact specific heat from just one of these states as well, which implies that \mbox{$C^{(q)}_v = C_v$} for this decomposition.
Formally, then, this particular basis is the most optimal one for sampling the average energy of the system. However, constructing even one
state $\ket{\Theta}$ amounts to diagonalizing the entire Hamiltonian, which would render a sampling approach irrelevant. 

In the end, then, the METTS decomposition achieves both a remarkably high sampling efficiency in practice, as quantified by \mbox{$C_v - C^{q}_v$}, 
while also being one of the least costly decompositions to produce.

\subsection{In What Sense are METTS Minimally Entangled?}

Consider a system consisting of just two spin 1/2 moments, $A$ and $B$ in a pure state $\ket{\psi}$. We call $\ket{\psi}$ entangled if it cannot be factorized, that is, if it is not a CPS. 

To compute the entanglement entropy $S[\psi]$ one first forms the density matrix $\rho_A$ describing spin $A$ alone
by tracing out the states of $B$
\begin{equation}
\rho_A = \text{Tr}_B \ket{\psi}\bra{\psi} \ .
\end{equation} 
One may define $\rho_B$ likewise. In terms of these reduced density matrices, $S[\psi]$ is then
\begin{equation}
S[\psi] = - \text{Tr}[\rho_A \log_2(\rho_A)] = - \text{Tr}[\rho_B \log_2(\rho_B)] \ . \label{eqn:EE}
\end{equation}

At finite temperature, however, $A$ and $B$ are not in a pure state $\ket{\psi}$, but in the mixed state described 
by the density matrix \mbox{$\rho = \frac{1}{\mathcal{Z}} e^{-\beta H}$}. While for a mixed state one can no longer define an entanglement entropy
in the same manner,
one could choose a particular decomposition of $\rho$ as in Eq.~(\ref{eqn:decomposition})
and compute its average entanglement entropy
\begin{equation}
S_\phi[\rho] = \sum_i \frac{P(i)}{\mathcal{Z}}\: S[\phi(i)] \ .
\end{equation}
However, $S_\phi[\rho]$ turns out to depend upon which decomposition we choose. 

\begin{figure}[htp]
\includegraphics[width=\columnwidth]{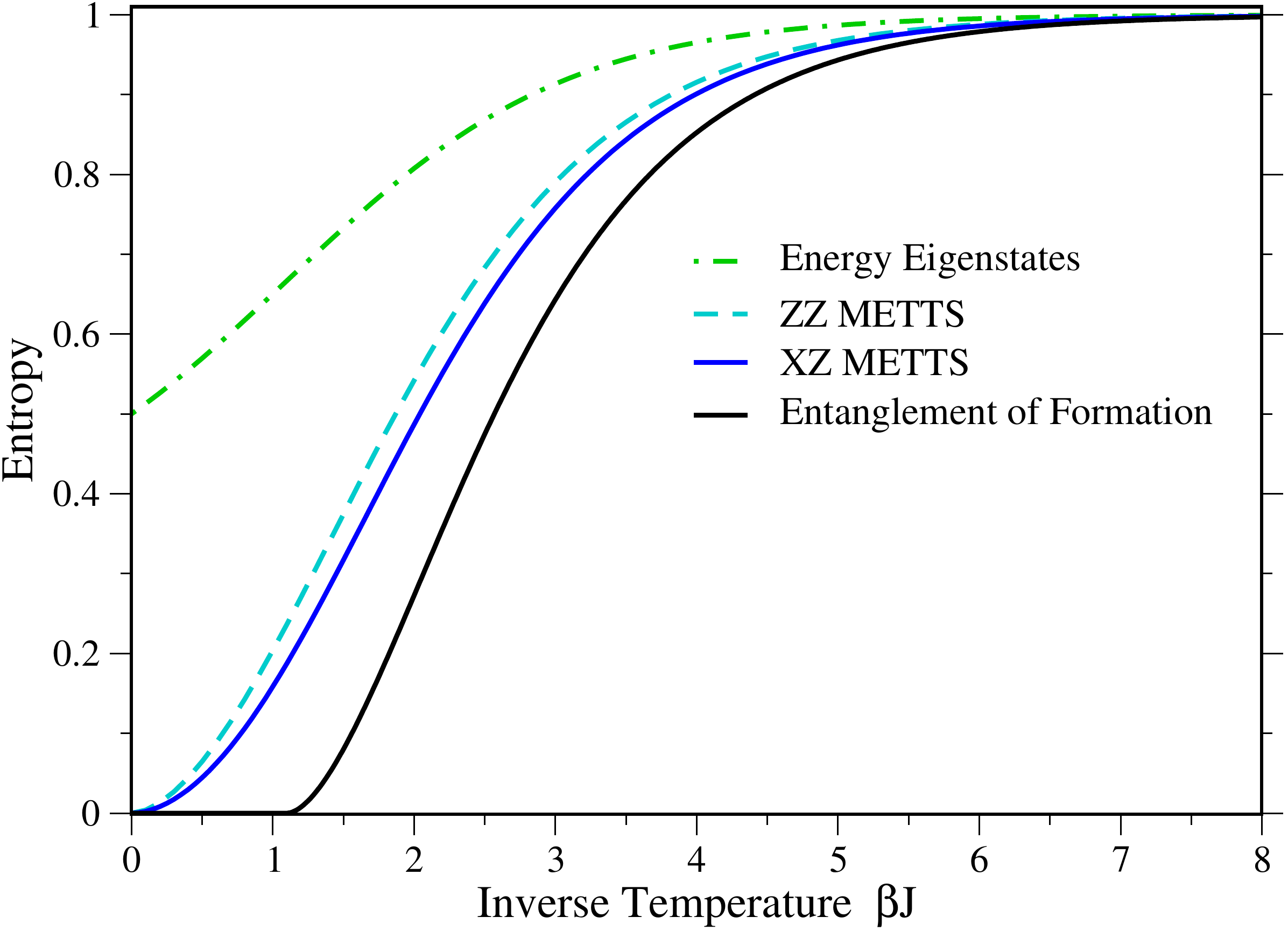}
\caption{The average entanglement of the ZZ and XZ METTS density matrix decompositions for the two site \mbox{$S= 1/2$} Heisenberg antiferromagnet. 
The XZ METTS form the optimal METTS decomposition of the system, but their average entropy exceeds the entanglement of formation $E[\rho]$. 
Also shown is the average entanglement of the energy eigenstate decomposition.}
\label{fig:AFM_entanglement}
\end{figure}

A more meaningful measure of mixed state entanglement 
is the \emph{entanglement of formation} $E[\rho]$ defined as the minimum value of $S_\phi[\rho]$ amongst all possible decompositions~\cite{bennett-96} 
\begin{equation}
E[\rho] = \text{min}_\phi \: S_\phi[\rho] \ . \label{eqn:EoF}
\end{equation}
While such a minimization would be difficult to compute in general, an exact expression has been 
found by Wootters for the case of a two qubit system.~\cite{wootters-98} So, for this simple system at least, we may unambiguously 
check whether the average entanglement entropy of the METTS decomposition saturates the lower bound set by $E[\rho]$. 
In other words, we may explicitly check if METTS are minimally entangled.

Before we make this comparison, though, let us review Wootters' method for computing the entanglement of formation. 
First define the concurrence of a pure state $\ket{\psi}$ of two qubits as
\begin{equation}
C[\psi] = |\bracket{\psi}{\tilde{\psi}}|
\end{equation}
where $\ket{\tilde{\psi}}$ is the time reversal conjugate of $\ket{\psi}$.
Concurrence is an alternative measure of entanglement, and may be used to compute
the standard von Neumann entanglement entropy Eq.~(\ref{eqn:EE}) through the relation
\begin{equation}
S[\psi] = f\left(\frac{1+\sqrt{1-(C[\psi])^2}}{2}\right)
\end{equation}
where $f(x) = -x \log_2(x) - (1-x)\log_2(1-x)$.

Now for a general mixed state of two qubits, it may be shown that there exists at least one optimal decomposition of $\rho$ which 
saturates the minimum in Eq.~(\ref{eqn:EoF}) and which consists of pure states 
all having the \emph{same} concurrence, and hence the same entanglement entropy.
This minimal concurrence value is given by
\begin{equation}
C[\rho] = \max\{0,\lambda_1-\lambda_2-\lambda_3-\lambda_4\}
\end{equation}
where the $\lambda_i$ are the square roots of the eigenvalues of the matrix $\rho\, (\sigma_y \otimes \sigma_y) \rho^{*} (\sigma_y \otimes \sigma_y)$.
It therefore follows that the entanglement of formation of a two qubit system is given by
\begin{equation}
E[\rho] = f\left(\frac{1+\sqrt{1-(C[\rho])^2}}{2}\right) \:.
\end{equation}

Equipped with this explicit formula for the entanglement of formation, let us now consider
a specific system, namely 
the two site $S= 1/2$ Heisenberg model
with Hamiltonian
\begin{equation}
H = J\: \vec{S}_A \cdot \vec{S}_B
\end{equation}
and take the antiferromagnetic $J > 0$ case. 

As \mbox{$\beta\rightarrow\infty$}, the unique ground state $\ket{\psi_0}$ of $H$ is the singlet formed out the the spins $A$ and $B$, which has an entanglement
entropy $S[\psi_0] = 1$. So, the entanglement of formation must also be $E[\rho] = 1$ at zero temperature.
In the opposite limit as $\beta$ goes to zero, $E[\rho]$ must go to zero. As shown in Fig.~\ref{fig:AFM_entanglement}, it actually reaches zero at a finite temperature of \mbox{$\beta J = \ln(3)$}. Thus for any smaller value of $\beta J$, the density matrix is separable and may be decomposed entirely in terms of CPS .~\cite{vedral-01}

Now let us compute the average entanglement of a METTS decomposition of this system. To do so, we must first choose
the underlying CPS $\ket{i}$ to be used in producing our METTS. One possible choice is to take the set $\ket{i}$ to be the complete basis of 
$\hat{z}$ eigenstates. Because this yields a decomposition consisting of only four METTS we can calculate them exactly. 
The resulting average entanglement entropy of this ``ZZ METTS'' ensemble is plotted in Fig.~\ref{fig:AFM_entanglement}. 

Note that we could have also used the pure state algorithm to 
sample this decomposition by collapsing each METTS into the $\hat{z}$ basis on every site and then the $\hat{x}$ basis on every site on alternate steps. The METTS produced after either type of collapse can then be used to measure the entanglement since the system is rotationally invariant. 
Collapsing into the $\hat{z}$ basis alone, on the other hand, will not sample all four METTS since the Hamiltonian conserves total $S^z$.

We may also choose CPS bases other than the $S^z$ eigenstates and thereby find METTS decompositions with even lower average entanglement. 
Consider the XZ METTS derived from the orthonormal CPS basis
\mbox{$\{\ket{\uparrow\rightarrow},\ket{\uparrow\leftarrow},\ket{\downarrow\rightarrow},\ket{\downarrow\leftarrow} \}$}, where up and down arrows denote $S^z$
eigenstates and left and right arrows denote $S^x$ eigenstates. As shown in Fig.~\ref{fig:AFM_entanglement}, the average entanglement of these METTS are
uniformly lower than the ZZ METTS. 

What is more, the variance of both energy and entropy is exactly zero for the XZ METTS ensemble. This follows from the fact that any XZ METTS can
be transformed into any other by global spin rotations that commute with $H$. Therefore we can conclude that the XZ METTS are the optimal METTS 
decomposition for this system, at least up to a global spin rotation. 
Note that one may also sample the XZ METTS with the pure state method by collapsing
them into the $\hat{x}$ and $\hat{z}$ bases on alternating steps \emph{and} on alternating sites. 

\begin{figure}[htp]
\includegraphics[width=\columnwidth]{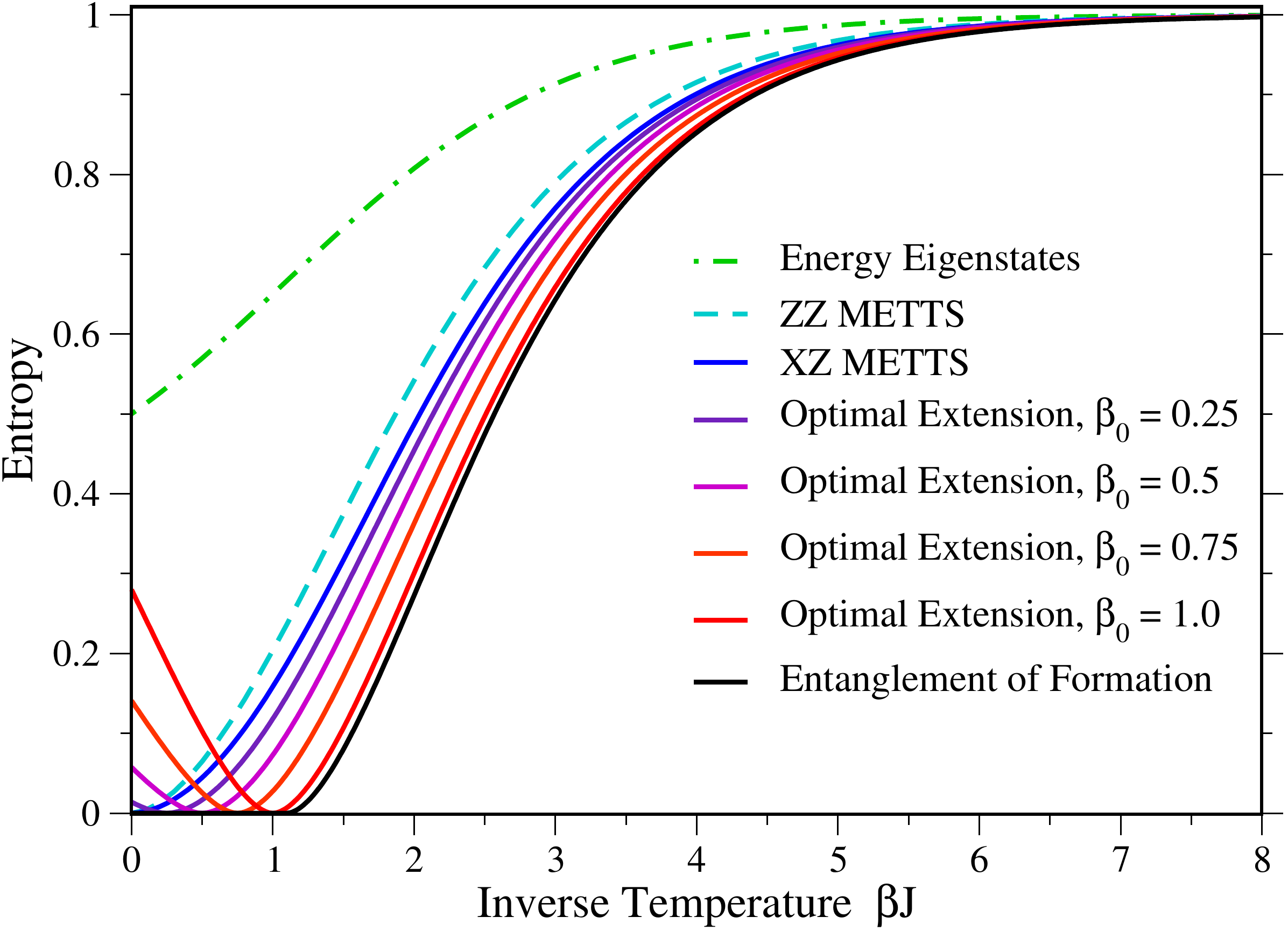}
\caption{The average entanglement entropy of extended decompositions of the two site $S=1/2$ Heisenberg antiferromagnet. 
Each optimal extension reaches the minimum allowed entanglement at some temperature $\beta_0$, but still exceeds the entanglement of the optimal XZ METTS 
at high enough temperatures. 
Also shown are the entanglement of the ZZ METTS decomposition and the energy eigenstate decomposition. 
}
\label{fig:AFM_entanglement_comp}
\end{figure}

Regardless of the basis we choose, however, one can see from Fig.~\ref{fig:AFM_entanglement} that the average entanglement of 
each METTS decomposition goes to zero
smoothly as \mbox{$\beta\rightarrow 0$} and always exceeds $E[\rho]$. Therefore, because even the optimal METTS decomposition 
is greater than $E[\rho]$ for any finite temperature, we can conclude that METTS are \emph{not} minimally entangled.

\begin{figure}[htp]
\includegraphics[width=\columnwidth]{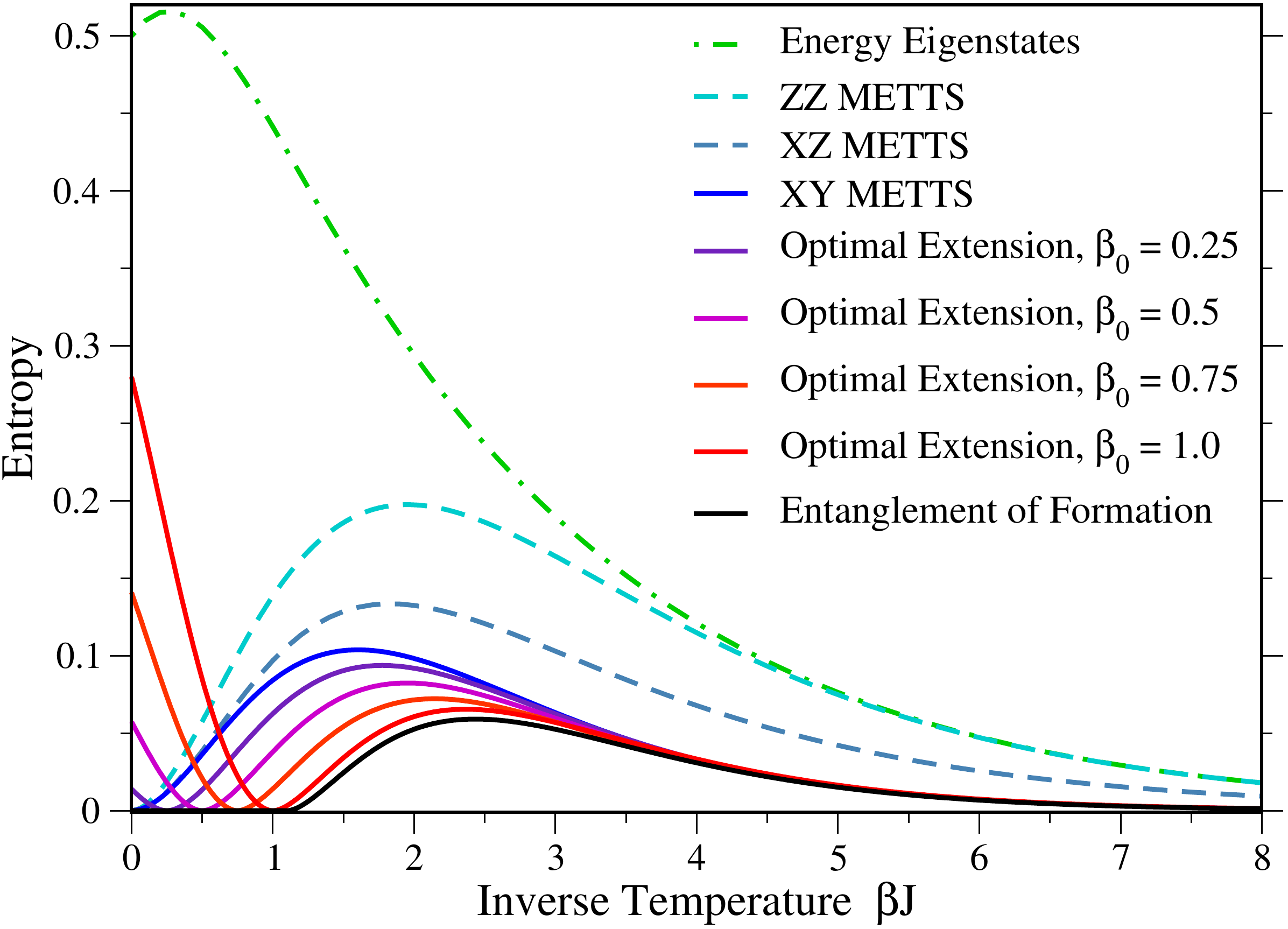}
\caption{The average entanglement entropy of extended decompositions of the two site $S=1/2$ Heisenberg antiferromagnet 
with an applied field $h=1.5 \,J$. In the presence of the field, the optimal METTS decomposition is given by the XY METTS. 
Also shown are the entanglement of the XZ and ZZ METTS decompositions and the energy eigenstate decomposition. 
}
\label{fig:hAFM_entanglement_comp}
\end{figure}

However, we may ask if the optimal METTS decomposition is instead minimally entangled in a more limited sense. 
For this purpose it will be helpful to look at a special class of thermal density matrix decompositions.
Beginning from a decomposition at some particular temperature $\beta_0$
\begin{equation}
\rho_0 = \frac{e^{-\beta_0 H}}{\mathcal{Z}_0} = \sum_i \,p_{i}\:\ket{\psi_i} \bra{\psi_i} \:, \label{eqn:arb_decomp}
\end{equation}
one can generate a decomposition for $\rho$ at a different temperature $\beta$ by making use of the relation
\begin{align}
\rho &  = \frac{e^{-\beta H}}{\mathcal{Z}} = \frac{\mathcal{Z}_0}{\mathcal{Z}}\: e^{-(\beta-\beta_0) H /2}\, \rho_0 \,e^{-(\beta - \beta_0) H /2} \nonumber \\
       &  = \sum_i \frac{p_i \mathcal{Z}_0}{\mathcal{Z}} Q(i)\: \ket{\psi_i(\beta)} \bra{\psi_i(\beta)} \label{eqn:extension}
\end{align}
where the pure states defining the new decomposition are given by
\begin{align}
\ket{\psi_i(\beta)} & =  e^{-(\beta-\beta_0) H /2} \ket{\psi_i} / Q(i)^{1/2} \nonumber \\
Q(i) & =  \bra{\psi_i} e^{-(\beta-\beta_0) H} \ket{\psi_i} \:.
\end{align}
We may consider this new decomposition the \emph{extension} of the decomposition of Eq.~(\ref{eqn:arb_decomp}), or 
an extended decomposition for short. 
Note that the entanglement of an extended decomposition is always an analytic function of $\beta$. 
For our two qubit system then, this implies that no extended decomposition can have an entanglement equal to 
$E[\rho]$ at all temperatures.

In fact, we would like to conjucture that no extended decomposition can even have an entanglement below the optimal
METTS decomposition at all temperatures. To see why this may be the case, let us attempt to construct a counterexample
for the two qubit system. At any temperature $\beta_0$, we can construct an optimal decomposition, following Wootters,
such that the entropy of each state $\ket{\psi_i}$ is equal to the entanglement of formation. Then, because the extensions of these decompositions 
will have the least possible entanglement at $\beta_0$, we expect that they will also have low entanglement at other temperatures.

Now, there can be multiple optimal decompositions for each starting temperature $\beta_0$. However, we have found that there is always 
one whose extension is less entangled than the others and has zero entropy variance for all temperatures, not just $\beta_0$. 
As shown in Fig.~\ref{fig:AFM_entanglement_comp}, while such optimal extensions have less entanglement than the optimal METTS at low temperatures,
they eventually exceed the optimal METTS as the temperature is increased. Moreover, the less entangled these extensions become at high temperature,
the more they resemble the optimal METTS ensemble itself. 

To check that this behavior is general, we can also study the two qubit system in a field with Hamiltonian
\begin{equation}
H = J\: \vec{S}_A \cdot \vec{S}_B - h\:(S^z_A + S^z_B) \:.
\end{equation}
Here we focus on the case \mbox{$J>0$} and \mbox{$h\geq0$}.
For small fields, the temperature dependence of the entanglement of formation $E[\rho]$ remains similar to the zero field case.
Then at $h=J$, the system undergoes a first-order quantum phase transition such that the ground state is no longer a maximally entangled singlet, but the 
fully polarized CPS with zero entanglement. And once $h$ is greater than $J$, the entanglement of formation no longer behaves as a
monotonic function of temperature. Regardless of the value of $h$, though, $E[\rho]$ is strictly zero for $\beta J \leq \ln(3)$ indicating that $\rho$ becomes separable above this temperature. 

In Fig.~\ref{fig:hAFM_entanglement_comp} we have taken a particular value of the field $h > J$ and plotted the entropies of various optimal extensions.
As before, each extension either exceeds the optimal METTS entanglement over a significant temperature range or else 
approaches the optimal METTS decomposition ever more closely as this range is reduced. Also, it is interesting to note that in the presence of a field 
it is the XY METTS (generated from $S^x$ and $S^y$ eigenstates) which now form the optimal METTS decomposition, at least up to a rotation about the $\hat{z}$ axis. This is because the XZ METTS are no longer required to be physically identical by the symmetries of the Hamiltonian and will exhibit a finite entropy variance.

Going forward, it will be interesting to explore the extent to which the observations made here hold for larger systems and more complex models. It would be especially useful to know if an optimal METTS decomposition with ideal sampling efficiency always exists and whether it can be efficiently computed. 

And although the average entanglement entropy of a METTS decomposition is not an entanglement measure in a strict sense (for instance, it is not zero when $\rho$ is separable),~\cite{vedral-97} it is conceivable that its scaling behavior could reveal non-trivial information about a system. Whether or not one must find 
an optimal METTS decomposition to access such information remains to be seen. 
But, because the METTS pure state algorithm is efficient enough to simulate moderately large two-dimensional systems, 
the entanglement properties of METTS could turn out to be a useful tool in characterizing exotic phases or critical points.

\section*{Acknowledgements}

EMS would like to thank Leon Balents for making this work possible and for providing support through \mbox{NSF Grant No.\ DMR-0804564}. 
EMS also acknowledges helpful discussions with Thomas Barthel and Ryan Mishmash, and thanks the 2009 ICTS Condensed Matter Programme for hospitality. 
SRW would like to acknowledge support from \mbox{NSF Grant No.\ DMR-0907500}. 

The quantum Monte Carlo calculations appearing in this work were performed using the ALPS looper application.\cite{alps} Thanks to Microsoft Station Q for providing the computing resources for these calculations.


\appendix

\section*{Appendix: Working with Matrix Product States and Matrix Product Operators \label{appendix:MPSandMPO} }

A matrix product state (MPS) is a scheme for representing a wavefunction, either exactly or approximately, by writing its amplitudes in some particular basis as a product of matrices. A general MPS may be written as
\begin{equation}
\ket{\psi_A} = \sum_{\{s\}} A^{s_1} A^{s_2}\cdot\cdot\cdot A^{s_N} \ket{s_1} \ket{s_2}\ldots\ket{s_N} \label{eqn:MPS}
\end{equation}
where the site labels $s_i$ take on values $s_i = 1,2,\ldots,d$ labeling the local basis.

For an adjacent pair of matrices $A^{s_i}_{\mu\nu} A^{s_{i+1}}_{\nu\lambda}$, if the index $\nu$ summed between the matrices takes on values $1,2,\ldots,m$ one says that the bond between matrices $i$ and $i+1$ has bond dimension $m$. Though $m$ can vary from bond to bond, in the following discussion we assume that it has a fixed value for a given MPS.
And, while in the presence of periodic boundary conditions one must trace over the above product of matrices in order to reduce it to a scalar amplitude, in what follows we work exclusively with open boundary conditions such that $A^{s_1}$ and $A^{s_N}$ are just $1\times m$ and $m\times1$ matrices, respectively.

A matrix product operator (MPO) is defined similarly to an MPS except that the decomposition is over a basis of local operators. A general MPO 
may therefore be written as
\begin{equation}
\op{W} = \sum_{\{t,s\}} W^{t_1s_1}\cdot\cdot\cdot W^{t_N s_N}\ket{t_1}\ldots\ket{t_N} \bra{s_1}\ldots\bra{s_N} \label{eqn:MPO}
\end{equation}
where for open boundary conditions the dimensions of the first and last matrix are again $1\times m$ and $m\times1$.

\subsection{Creating MPOs}

The simplest MPOs to create are those describing products of local operators, that is \mbox{$\op{W} = \Pi_j \hat{\mathcal{O}}_j$}. In particular, one can represent a single site operator $\hat{\mathcal{O}}_i$
this way by taking \mbox{$\hat{\mathcal{O}}_j = 1$} for all $j\neq i$. If the operator $\op{W}$ is such a site-factorized product, its MPO representation has a simple product structure as well. Each $W^{t_is_i}$ is actually just a $1\times1$ matrix defined by
$W^{t_is_i} = \bra{t_i}\hat{\mathcal{O}}_i\ket{s_i}$.

Another simple class of MPOs are those describing translationally invariant sums of operators in one dimension in which case the matrices 
$W$ can be taken to be identical (except at the boundaries) and of lower triangular form.
For example, the MPO of the operator \mbox{$S^z_{\mbox{\tiny tot}} = \sum_{i=1}^{N} S^z_i$} may be represented in terms of the bulk matrices
\[
W^{t_is_i} = \left[ \begin{array}{cc} \delta^{t_is_i} & 0 \\ (S^{z}_i)^{t_is_i} & \delta^{t_is_i} \end{array} \right]
\]
together with the boundary matrices \mbox{$W^{t_1s_1} = \left[0\ \delta^{t_1s_1}\right]$} and
\mbox{$W^{t_Ns_N} = \left[\delta^{t_Ns_N}\ 0\right]^{T}$}. Translationally invariant one dimensional Hamiltonians can likewise be represented
by translationally invariant lower triangular MPOs of bond dimension $3$ and higher.\cite{mcculloch-07} 

However, there may not be always such a simple, explicit 
prescription for operators of interest, in particular Hamiltonians of two dimensional systems.
For use in such cases, we present below general algorithms for the addition and multiplication of arbitrary MPOs.

Equipped with these algorithms, the MPO representation of any Hamiltonian that is a sum of local terms may be found by first obtaining the MPOs
for each term as described above and then summing them together. The MPO multiplication algorithm is then useful for forming 
other important operators such as $H^2$ for specific heat measurements or $e^{-\tau H}$ for imaginary time evolution.

\subsection{Adding MPS and MPOs \label{appendix:adding}}

First consider adding two MPS $\ket{\psi_A}$ and $\ket{\psi_B}$ with bond dimensions $m_A$ and $m_B$. 
If $\ket{\psi_C} = \ket{\psi_A} + \ket{\psi_B}$,
then
\begin{equation}
C^{s_i} = A^{s_i} \oplus B^{s_i} \ . 
\end{equation}
which in block form is 
\begin{equation} 
C^{s_i} = \left[ \begin{array}{ccc}
A^{s_i} & 0  \\
0 & B^{s_i}  
\end{array} \right] \:.
\end{equation} 
More formally, one may write this definition of $C^{s_i}$ as
\begin{multline}
C^{s_i}_{\chi_{i-1}\chi_i} = \!\!\!\!\! \sum_{\alpha_{i-1},\alpha_i,\beta_{i-1},\beta_i} \!\! \delta_{\chi_{i-1}\alpha_{i-1}} A^{s_i}_{\alpha_{i-1}\alpha_i} \delta_{\alpha_i\chi_i} \\
\mbox{} +  \delta_{\chi_{i-1}(\beta_{i-1}+m_A)} B^{s_i}_{\beta_{i-1}\beta_i} \delta_{(\beta_i+m_A)\chi_i} \label{eqn:directsum}
\end{multline}
where the $\chi$ indices run over values $1,2,\ldots,(m_A+m_B)$.

After constructing the $C$ matrices as above, the addition has been carried out exactly. However, one usually wants to work with an MPS for $\ket{\psi_C}$ that
has a bond dimension comparable to $m_A$ or $m_B$, not $m_A+m_B$. One therefore usually truncates the MPS representation for $\ket{\psi_C}$ as follows.

Treating $C_1 = C^{s_1}_{\chi_1}$ as a matrix $C_{s_1\chi_1}$, compute its SVD 
\begin{equation}
C_{s_1\chi_1} = U_{s_1\chi^\prime_1} \Lambda_{\chi^\prime_1} V^{\dagger}_{\chi^\prime_1\chi_1} \ .
\end{equation}
During this step, only $m_C$ singular values are kept such that $\chi^\prime_1 = 1,2,\ldots,m_C$. 
Finally, replace $C_1$ with $C^{s_1}_{\chi^\prime_1} = U_{s_1\chi^\prime_1}$ and replace $C_2$ with 
$C^{s_2}_{\chi^\prime_1\chi_2} = \Lambda_{\chi^\prime_1} V^{\dagger}_{\chi^\prime_1\chi_1} C^{s_2}_{\chi_1\chi_2}$. 

The next bond in the truncated MPS for $\ket{\psi_C}$ is now found by treating the new $C_2$ as a matrix $C_{(s_2\chi^\prime_1)\chi_2}$ and 
computing its SVD. The matrix $C_2$ is then set to $U_2$ and $C_3$ is multiplied by $\Lambda_2V^\dagger_2$ just as $C_2$ was in the first step. 
This truncation process may then be likewise repeated for every remaining $C$ matrix. 

The addition of MPOs proceeds in complete analogy to the addition of MPS. To compute the MPO $\op{Z}$ such that 
\mbox{$\op{Z} = \op{W} + \op{X}$}, one defines the matrices $Z$ as 
\begin{equation}
Z^{s_it_i} = W^{s_it_i} \oplus X^{s_it_i}
\end{equation}
where the only difference from the MPS case is that there are now $d^2$ matrices to be added instead of $d$. Thinking 
of the pair of indices $s_i$ and $t_i$ as a fat index $(s_it_i)$, one may likewise truncate the matrices $Z$ by performing 
sequential SVDs as above. 


The MPS addition algorithm above has a computational cost of $N m^3 d^2$ for \mbox{$m_A = m_B = m_C = m$}. 
Similarly, the addition of MPOs scales as $N k^3 d^4$ where $k$ is the MPO bond dimension.

\subsection{Multiplying MPOs}

Consider the task of multiplying two MPOs $\op{W}$ and $\op{X}$ with the goal of producing a product MPO \mbox{$\op{Z} = \op{W}\op{X}$}. Let these MPOs be written as products of $W$, $X$ and $Z$ matrices with bond dimension $k$. 

This multiplication may be carried out one site at a time in a process analogous to the zip-up algorithm discussed above. Therefore, the first 
step is to map each MPO to an MPS by combining its $t$ and $s$ (or primed and unprimed) indices at each site into a fat index. These MPS should then 
be transformed such that their orthogonality center is at the first site, and then mapped back into MPOs. In doing so, one ensures that
in the truncation step below, the basis of states surrounding the local tensor is not overly ill-conditioned. However, 
because this basis will not be orthonormal, it is important to use a truncation error cutoff instead of a fixed $k$ to determine the final MPO bond dimension. 

Now, to carry out the multiplication beginning with the first site, define the tensor $R_1$ as
\begin{equation}
R^{s_1t_1}_{\omega_1\xi_1} = \sum_{u_1} W_{\omega_1}^{t_1u_1} X_{\xi_1}^{u_1s_1} \ .
\end{equation}

Thinking of $R_1$ as a matrix $R_{(s_1t_1)(\omega_1\xi_1)}$, compute its SVD to obtain
\begin{equation}
R_{(s_1t_1)(\omega_1\xi_1)} = U_{(s_1t_1)\zeta_1} \Lambda_{\zeta_1} V^{\dagger}_{\zeta_1(\omega_1\xi_1)}\ ,
\end{equation}
where the singular values are truncated such that only the $k$ largest are kept. The first matrix of $\op{Z}$ may now be defined as
$Z_{\zeta_1}^{s_1t_1} = U_{(s_1t_1)\zeta_1}$.

The remaining $Z$ matrices may now be found by proceeding recursively. Define $R_2$ in terms of $R_1$ as
\begin{equation}
R^{s_2t_2}_{\zeta_1\omega_2\xi_2} = \! \sum_{s_1t_1} U^{\dagger}_{\zeta_1(s_1t_1)} \left[ \sum_{\xi_1 u_2} \left[ \sum_{\omega_1} R^{s_1t_1}_{\omega_1\xi_1} W^{t_2u_2}_{\omega_1\omega_2} \right] X^{u_2s_2}_{\xi_1\xi_2} \right] \: . \label{eqn:Rtensor}
\end{equation}
Again, $R_2$ may be thought of as a matrix and its SVD computed as
\begin{equation}
R_{(s_2t_2\zeta_1)(\omega_2\xi_2)} = U_{(s_2t_2\zeta_1)\zeta_2} \Lambda_{\zeta_2} V^{\dagger}_{\zeta_2(\omega_2\xi_2)} \ .
\end{equation}
The second matrix of $\op{Z}$ may now be defined as \mbox{$Z^{s_2t_2}_{\zeta_1\zeta_2} = U_{(s_2t_2\zeta_1)\zeta_2}$} and the algorithm
continued by defining $R_3$ analagously to $R_2$ above in terms of $U_2$, $R_2$, $W_3$ and $X_3$.

If the contractions in Eq.~(\ref{eqn:Rtensor}) are carried out in the order indicated, this algorithm for multiplying MPOs scales as $N k^4 d^3$ with the most expensive step being the construction of the $R$ tensors.

\bibliography{METTS_Paper}

\end{document}